\documentclass[useAMS,usenatbib,usegraphicx]{mn2e}

\usepackage{amssymb} 
\usepackage{textcomp} 
\usepackage{appendix}
\usepackage{amsmath} 
\usepackage{fixltx2e} 
\usepackage{multirow} 
\usepackage{enumerate} 
\usepackage{mathptmx} 
\usepackage{lscape}  
\usepackage[colorlinks=true,citecolor=blue,linkcolor=blue,urlcolor=blue]{hyperref} 

\renewcommand{\th}{$^{\rm th}$}

\newcommand{\kms}{\,\rm km\ s^{-1}}
\newcommand{\ergs}{\,\rm erg\ s^{-1}}

\newcommand{\mic}{\,\mbox{$\mu$m}}
\newcommand{\Hz}{\,\rm{Hz}}
\newcommand{\kev}{\,\rm keV}

\let\AAold\AA
\renewcommand{\AA}{\text{\AAold}}
\newcommand{\eV}{\,{\rm eV}}
\newcommand{\cm}{\,{\rm cm}}
\newcommand{\pc}{\,{\rm pc}}
\newcommand{\kpc}{\,{\rm kpc}}

\newcommand{\ryd}{\,{\rm Ryd}}
\newcommand{\s}{\,{\rm s}}

\newcommand{\K}{\,{\rm K}}
\newcommand{\msun}{\,{\rm M_{\odot}}}
\newcommand{\zsun}{\,{\rm Z_{\odot}}}

\renewcommand{\mp}{m_{\rm p}}
\newcommand{\kB}{K_{\rm B}}
\newcommand{\sth}{\sigma_{\rm Th}}

\newcommand{\hnu}{\langle h\nu\rangle}
\newcommand{\ncrit}{n_{\rm crit}}

\newcommand{\lbol}{L_{\rm bol}}
\newcommand{\lledd}{L / L_{\rm{Edd}}}
\newcommand{\mbh}{M_{\rm BH}}

\newcommand{\aeuv}{\alpha_{\rm ion}}

\newcommand{\sV}{\sigma_{\rm *}}

\newcommand{\hi}{\text{H~{\sc i}}}
\newcommand{\hii}{\text{H~{\sc ii}}}
\newcommand{\nii}{\text{[N~{\sc ii}]}}

\newcommand{\sii}{\text{[S~{\sc ii}]}}
\newcommand{\six}{\text{S~{\sc ix}]}}

\newcommand{\oi}{\text{[O~{\sc i}]}}
\newcommand{\oii}{\text{[O~{\sc ii}]}}
\newcommand{\oiii}{\text{[O~{\sc iii}]}}
\newcommand{\oiv}{\text{[O~{\sc iv}]}}
\newcommand{\ovip}{\text{O~{\sc vi}}}

\newcommand{\fex}{\text{Fe~{\sc x}}}
\newcommand{\caii}{\text{[Ca~{\sc ii}]}}

\newcommand{\neiii}{\text{[Ne~{\sc iii}]}}
\newcommand{\nev}{\text{[Ne~{\sc v}]}}
\newcommand{\nevi}{\text{[Ne~{\sc vi}]}}
\newcommand{\neviiip}{\text{Ne~{\sc viii}}}
\newcommand{\fevii}{\text{[Fe~{\sc vii}]}}
\newcommand{\mgiip}{\text{Mg~{\sc ii}}}
\newcommand{\civp}{\text{C~{\sc iv}}}

\newcommand{\Ha}{\text{H$\alpha$}}

\newcommand{\Hb}{\text{H$\beta$}}

\newcommand{\Heii}{He~{\sc ii}}

\newcommand{\aap}{A\&A}
\newcommand{\araa}{ARA\&A}
\newcommand{\apjl}{ApJ}
\newcommand{\apjs}{ApJS}
\newcommand{\apj}{ApJ}
\newcommand{\aj}{AJ}
\newcommand{\mnras}{MNRAS}

\newcommand{\pasp}{PASP}

\newcommand{\lbha}{L_{\rmn{bH\alpha}}}

\newcommand{\lnha}{L_{\rmn{nH\alpha}}}

\newcommand{\loiii}{L_{\oiii}}

\newcommand{\lx}{L_{\rm X}}

\newcommand{\dv}{{v}}

\newcommand{\fion}{f_{\hi}}

\newcommand{\lion}{L_{\rm ion}}
\renewcommand{\ng}{n_{\gamma}}

\newcommand{\rgi}{r_{\rm inf}}

\newcommand{\lp}{l_{\rm P}}

\newcommand{\cloudy}{{\sc Cloudy}}

\newcommand{\HST}{{\it HST}}
\newcommand{\Chandra}{{\it Chandra}}
\newcommand{\IUE}{{\it IUE}}
 
\title[Radiation Pressure Confinement]{Radiation Pressure Confinement - I. Ionized Gas in the ISM of AGN Hosts}
\author[Jonathan Stern, Ari Laor and Alexei Baskin]
{Jonathan Stern,\thanks{E-mail: \href{mailto:stern@physics.technion.ac.il} {stern@physics.technion.ac.il}  (JS);
                        \href{mailto:laor@physics.technion.ac.il}  {laor@physics.technion.ac.il}   (AL);
                        \href{mailto:alexei@physics.technion.ac.il}{alexei@physics.technion.ac.il} (AB) }
 Ari Laor and Alexei Baskin \\
 Department of Physics, Technion -- Israel Institute of Technology, Haifa 32000, Israel}

\begin{document}
\maketitle

\begin{abstract}
We analyze the hydrostatic effect of AGN radiation pressure on optically thick gas in the host galaxy.
We show that in luminous AGN, the radiation pressure likely confines the ionized layer of the illuminated gas.
Radiation pressure confinement (RPC) has two main implications. 
First, the gas density near the ionization front is $7\times10^4~L_{\rm i, 45} r_{50}^{-2}\cm^{-3}$, where $L_{\rm i,45}$ is the ionizing luminosity in units of $10^{45}\ergs$ and $r_{50}$ is the distance of the gas from the nucleus in units of $50\pc$.
Second, as shown by Dopita et al., the solution of the ionization structure within each slab is unique, independent of the ambient pressure.
We show that the RPC density vs. distance relation is observed over a dynamical range of $\sim10^4$ in distance, from sub-pc to kpc from the nucleus, and a range of $\sim10^8$ in gas density, from $10^3$ to $10^{11}\cm^{-3}$. 
This relation implies that the radiative force of luminous AGN can compress giant molecular clouds in the host galaxy, and possibly affect the star formation rate. 
The unique ionization structure in RPC includes a highly ionized X-ray emitting surface, an intermediate layer which emits coronal lines, and a lower ionization inner layer which emits optical lines. 
This structure can explain the observed overlap of the extended X-ray and optical narrow line emission in nearby AGN. 
We further support RPC by comparing the predicted ratios of the narrow lines strength and narrow line widths with available observations.
We suggest a new method, based on the narrow line widths, to estimate the black hole mass of low luminosity AGN.
\end{abstract}


\section{Introduction}

Observations of emission lines in active galaxies point to the presence of photoionized gas in a wide range of radii, ionization states, gas densities and velocities. 
The radii $r$, H nuclei densities $n$, and velocities $v$ seem to be strongly coupled, with dense $n\sim10^{9-11}\cm^{-3}$ and fast $v\sim3\,000\kms$ ionized gas appearing on sub-pc scales (the broad line region, or BLR), while lower $v\sim300\kms$ and lower $n\sim10^{2-5}\cm^{-3}$ ionized gas appears at scales of tens of parsecs to several kiloparsecs (the narrow line region, or NLR). In some low luminosity active galactic nuclei (AGN), an intermediate region with $v\sim1\,000\kms$ and $n\sim10^{6-8}\cm^{-3}$ is also observed  (\citealt{FilippenkoHalpern84}; \citealt{Filippenko85}; \citealt{AppenzellerOestreicher88}; \citealt{Ho+96}). 
This decrease of $n$ and $v$ with increasing $r$ seems to appear also within specific regions. Resolved observations of the NLR show $n$ increases towards the nucleus (\citealt{Kraemer+00}; \citealt{Barth+01}; \citealt{Bennert+06a, Bennert+06b}; \citealt{Walsh+08}; \citealt{Stoklasova+09}), while the unresolved intermediate line region shows an increase of $n$ with $v$ (see \citealt{FilippenkoHalpern84} and citations thereafter). 
An association of $v$ with $r$ is expected if the gas kinematics near the center are dominated by the black hole gravity, but what causes the association of $n$ with $r$?

On the other hand, the large scale stratification seen in $n$ is not observed in the ionization state. 
Quite the contrary is true -- in both the NLR and the BLR ions from a wide range of ionization potentials (IP) are commonly observed, including narrow lines of \sii\ (IP=10\eV), \oiii\ (IP=35\eV), \nev\ (97\eV), \fex\ (234\eV), and broad lines of \mgiip\ (8\eV), \civp\ (48\eV), \ovip\  (114\eV) and \neviiip\ (207\eV). 
Moreover, resolved maps of emission lines with widely different IP indicate that the high ionization gas and low ionization gas are co-spatial. A strong spatial correlation is seen between the high IP extended X-ray line emission and the relatively low IP \oiii\ emission (\citealt{Young+01}; \citealt{Bianchi+06}; \citealt{Massaro+09}; \citealt{Dadina+10}; \citealt{Balmaverde+12}). 
A similar correlation is also seen between the optical and near infrared high IP emission and the \oiii\ emission (\citealt{Mazzalay+10, Mazzalay+13}). 
The co-spatiality of the high ionization and low ionization gas suggests a common origin of these two components. 

Also, in mid infrared emission lines, where extinction effects are minimal, high IP and low IP lines exhibit a very small dispersion in their luminosity ratios. The $\nev~14.32\mic$, $\oiv~25.89\mic$ and $\neiii~15.55\mic$ emission lines have IPs of 97\eV, 55\eV\ and 41\eV\ respectively, but the dispersion in their luminosity ratios between different objects is $\lesssim0.2$ dex (\citealt{Gorjian+07}; \citealt{Melendez+08}; \citealt{Weaver+10}). This small dispersion also suggests a common origin for the low ionization and high ionization gas. Why is low ionization gas always accompanied by high ionization gas, and vice-versa?

A possible physical source of the characteristics mentioned above is the mechanism which confines the ionized layer of the illuminated gas. 
On the back side, beyond the ionization front, cool dense gas can supply the confinement. However, an optically thin confining mechanism is required at the illuminated surface. Several optically thin confining mechanisms have been suggested for the ionized gas in AGN, usually for a specific region. 
A hot low $n$ medium in pressure equilibrium with cooler line-emitting gas has been proposed for the BLR (\citealt{Krolik+81}; \citealt{MathewsFerland87}; \citealt{Begelman+91}), and for the NLR (\citealt{KrolikVrtilek84}). 
Other proposed confining mechanisms include a low density wind striking the face of the gas (e.g. \citealt{WhittleSaslaw86}), and a magnetic field permeating the intercloud medium  (\citealt{Rees87} and  \citealt{Emmering+92} for the BLR; \citealt{deBruynWilson78} for the NLR). 
Most of the above suggestions require an additional component for confining the gas, implying that the gas pressure is an independent free parameter. 

However, one source of confinement is inevitable in a hydrostatic solution, and incurs no additional free parameters. Photoionization must be associated with momentum transfer from the radiation to the gas. Thus, the pressure of the incident radiation itself can confine the ionized layer of the illuminated gas, without requiring any additional components. In this simpler scenario, where the gas is radiation pressure confined (RPC), the gas pressure is set by the flux of the incident radiation.

\citeauthor{Dopita+02} (2002, hereafter D02), \cite{Groves+04a}, and \citeauthor{Groves+04b} (2004b, hereafter G04) showed that the gas pressure in the NLR gas is likely dominated by radiation pressure. They derived a slab structure where the ionization decreases with depth, which implies a common source for low IP and high IP emission lines. Also, this slab structure implies that the low ionization layer sees an absorbed spectrum, as observed in some nearby AGN 
(\citealt{Kraemer+00, Kraemer+09}; \citealt{Collins+09}). 
Building on the work of D02 and G04, we show that in RPC the same slab of gas which emits the low ionization emission lines can have a highly ionized surface which emits X-rays lines. We show that because the gas pressure is not an independent parameter, this slab structure is unique over a large range of $r$ and other model parameters. This specific structure is likely responsible for the tight relation between the low IP emission lines and the high IP lines observed in the X-ray, optical and IR.

If the ionized gas is RPC, then the pressure at the ionization front, where most of the ionizing radiation is absorbed, equals the incident radiation pressure, which is $\propto r^{-2}$. Since the temperature near the ionization front is $\sim10^4\K$, RPC implies $n \propto r^{-2}$. We show below that this $n \propto r^{-2}$ relation quantitatively reproduces the decrease of $n$ with $r$ seen in resolved observations of the NLR, and the increase of $n$ with $v$ seen in the unresolved intermediate line region. In a companion paper (\citealt{Baskin+13}, hereafter Paper II) we show that RPC also reproduces $n$ at the BLR. Together, these findings imply that RPC sets $n$ of ionized gas in active galaxies over a dynamical range of $\sim10^4$ in $r$, from sub-pc to kpc scale, and a dynamical range of $\sim10^8$ in $n$, from $10^3$ to $10^{11}\cm^{-3}$.

Hydrostatic radiation pressure effects were also applied to models of ionized gas in star forming regions (\citealt{Pellegrini+07,Pellegrini+11}; \citealt{Draine11a}; \citealt{Yeh+13}; \citealt{Verdolini+13}), and to models of `warm absorbers', i.e. ionized gas in AGN detected in absorption (\citealt{Rozanska+06}; \citealt{Chevallier+07}; \citealt{Goncalves+07}). 

The paper is built as follows.
In \S2.1 we present the necessary conditions for RPC. In \S2.2 -- \S2.6, we derive several analytical results from these conditions. We use \cloudy\ (\citealt{Ferland+98}) to carry out detailed numerical calculations. The emission line emissivities vs. $r$ implied by RPC are presented in \S2.7.
In \S3 we compare the RPC calculations with available observations. 
In \S4, we analyze the observational and theoretical evidence for the existence of dust in the ionized gas, which has a strong effect on the structure of RPC slabs. 
We discuss our results and their implications in \S5, and conclude in \S6.

\section{Radiation Pressure Confinement}\label{sec: rpc}

\newcommand{\mcl}{M_{\rm GMC}}
\newcommand{\dcl}{d_{\rm GMC}}
\newcommand{\prad}{P_{\rm rad}}
\newcommand{\pgas}{P_{\rm gas}}
\newcommand{\pgasz}{P_{\rm gas,0}}
\newcommand{\pgasf}{P_{{\rm gas,f}}}
\newcommand{\frad}{F_{\rm rad}}
\newcommand{\sigbar}{{\bar \sigma}}
\newcommand{\taubar}{{\bar \tau}}

\subsection{Conditions for RPC}\label{sec: rpc conditions}

We assume a one dimensional, hydrostatic, semi-infinite slab of gas.
The slab is assumed to be moving freely in the local gravitational field, so external gravity is canceled in the slab frame of reference. 
The ionized layer of this slab is confined by radiation pressure if it satisfies two conditions. 
The first condition is that the force applied by the radiation needs to be the strongest force applied to the gas. 
Under this condition, the hydrostatic equation is
\begin{equation}\label{eq: hydrostatic only radiation}
\frac{d\pgas}{dx} = \beta\frac{\frad}{c}n\sigbar
\end{equation}
where $\pgas$ is the gas pressure, $x$ is the depth into the slab measured from the illuminated surface, $\frad$ is the flux of ionizing radiation at $x$, and $\sigbar$ is the sum of the mean absorption and scattering opacity per H nucleus, weighted by the ionizing flux. We define a correction factor $\beta$, which accounts for the additional radiative force due to the absorption of non-ionizing photons in the ionized layer, and the correction due to anisotropic scattering. 
We show below that for a typical AGN spectral energy distribution (SED), $\beta\sim1$ in dust-less gas and $\beta\sim 2$ in dusty gas.
Other sources of pressure, including magnetic pressure and the pressure of the trapped emitted radiation, are assumed to be small compared to $\pgas$. 

The second condition, presented by D02, is that the radiation pressure needs to be significantly larger than the ambient pressure, i.e.,
\begin{equation} 
\beta\prad  >>  \pgasz 
\label{eq: prad>pgas}
\end{equation}
where $\prad$ is the pressure of the ionizing radiation at the illuminated surface ($\prad = \lion/4 \pi r^2 c$, $\lion$ is the luminosity at $1-1000\ryd$). 
For properties which are defined as a function of $x$, such as $\pgas$, we use a subscript `0' to denote a value of a certain variable at the illuminated surface, and a subscript `f' to denote a value near the `ionization front' -- the boundary between the \hii\ and \hi\ layers. 

At the ionization front, most of the ionizing radiation has been absorbed, therefore equations \ref{eq: hydrostatic only radiation} and \ref{eq: prad>pgas} imply that 
\begin{equation}
\pgasf = \pgasz + \beta\prad ~~~ \approx ~~~ \beta\prad 
\label{eq: pgasf}
\end{equation}
where we assumed that $x_{\rm f} << r$, so $\prad$ is not geometrically diluted with increasing $x$ (see \S 2.6). 

The natural units to discuss RPC is $\Xi \equiv \beta\prad/\pgas$. This definition of $\Xi$ differs from the common definition (\citealt{Krolik+81}) by a factor of $\beta$, which is the natural extension for including the effect of pressure from non-ionizing radiation. Also, we follow D02 and drop the factor of 2.3 in the original definition. 
In these units, equations \ref{eq: prad>pgas} and \ref{eq: pgasf} are simply
\begin{equation}
 \Xi_0 >> 1
\label{eq: min Xi}
\end{equation}
and 
\begin{equation}
 \Xi_{\rm f} = 1
\label{eq: Xi_f}
\end{equation}

\subsection{The $n$ near the ionization front}\label{sec: n vs r}
\newcommand{\betan}{\beta_{2}}

Near the ionization front, $T_{\rm f}\approx10^4\K$ to within a factor of $\lesssim 2$ (e.g. \citealt{Krolik99}). Therefore, using $\pgasf = 2.3n_{\rm f}\kB T_{\rm f}$ and $\beta=2$, eq. \ref{eq: pgasf} implies 
\begin{equation}\label{eq: n vs r}
n_{\rm f} = 7.4 \times 10^4 ~\frac{L_{{\rm i,} 45}}{r_{50}^2} ~ T^{-1}_{f,\ 4}~ \cm^{-3} 
\end{equation}
where $\lion = 10^{45}~L_{{\rm i,} 45}\ergs$, $r=50~r_{50}\pc$ and $T_{\rm f} = 10^4T_{f,4}\K$. Note that $n_{\rm f}$ is independent of $n_0$. 

\begin{figure}
\includegraphics{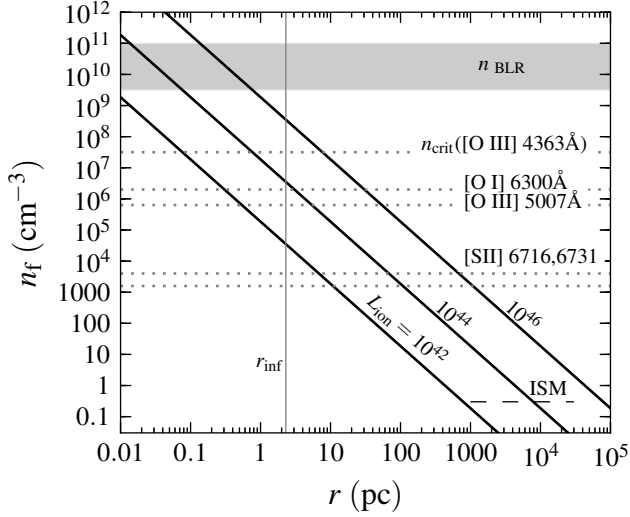}
\caption{The gas density near the ionization front implied by RPC, versus the distance from the AGN and the AGN luminosity.
Thick solid lines show the $n_{\rm f} \propto r^{-2}$ relation (eq. \ref{eq: n vs r}) for different $\lion$ (noted in $\ergs$). For comparison, we plot the typical $n$ of the BLR (gray stripe), the $\ncrit$ of various forbidden lines (dotted lines), the gravitational radius of influence for $\mbh=10^8\msun$ (solid gray, see \S\ref{sec: low lledd profiles}), and the typical ISM pressure in the solar neighborhood, in units of $T_4\cm^{-3}$ (short dashed line).
The AGN radiation pressure is larger than the typical ISM pressure at $r<25~L_{{\rm i,} 45}^{1/2}\kpc$. Since forbidden line emission drops at $n>\ncrit$, RPC implies a stratification of line emission according to $\ncrit$. In low luminosity AGN, high $\ncrit$ lines will be emitted from $r<\rgi$, and will have a wider profile than expected from $\sV$. 
}
\label{fig: n vs r}
\end{figure}

Equation \ref{eq: n vs r} is plotted in Figure \ref{fig: n vs r} for different $\lion$, assuming $T_{f,\ 4} = 1$. 
For comparison, we plot the typical ISM pressure in the solar neighborhood, $nT_4 = 0.3 \cm^{-3}$ (\citealt{Draine11b}). The pressure induced by the AGN radiation is stronger than the ISM pressure at $r < 25~L_{{\rm i,} 45}^{1/2}\kpc$. Therefore, the radiation pressure of Seyferts will likely have a significant effect on the ISM of the whole host galaxy, while Quasars can also significantly affect the pressure equilibrium in the circum-galactic medium. 

Fig. \ref{fig: n vs r} also shows the critical densities $\ncrit$ of various forbidden lines with relatively low ionization levels. For a broad distribution of $n$, the emission of a certain forbidden line is expected to peak at gas with $n\sim\ncrit$ (but see
a refinement in \S\ref{sec: emissivity vs. r}). 
The low ionization levels of these lines ensure they are emitted near the ionization front, so $n_{\rm f}$ is a measure of $n$ where these lines are emitted. 
Therefore, eq. \ref{eq: n vs r} implies that the forbidden line emission peaks at $r\sim14~(L_{{\rm i,} 45}/n_{\rm crit, 6})^{1/2}\pc$, where $\ncrit = 10^6 n_{\rm crit, 6}$. For a specific $\lion$ and forbidden line, the radius of peak emission can be read from the intersection of the appropriate solid and dotted lines in Fig. \ref{fig: n vs r}. Hence, RPC implies that the emission of forbidden lines should be stratified in $r$ according to their $\ncrit$. For example, the high $\ncrit$ \oiii\ $\lambda 4363$ line will be emitted from gas with $r$ which is 100 times smaller than the the gas which emits the low $\ncrit$ \sii\ doublet. In \S\ref{sec: observations}, we compare eq. \ref{eq: n vs r} with narrow line observations.

Also shown in Fig. \ref{fig: n vs r} is the range of $n$ observed in the BLR, $n_{\rm BLR} = 10^{9.5 - 11}\cm^{-3}$ (\citealt{DavidsonNetzer79}; \citealt{Rees+89}; \citealt{Ferland+92}; \citealt{Marziani+96}). The BLR is within the dust sublimation radius (\citealt{NetzerLaor93}; \citealt{Suganuma+06}), so $\beta({\rm BLR})\sim1$, lower by a factor of two than assumed in Fig. \ref{fig: n vs r}. 

\subsection{The effective $U$ near the ionization front}

D02 showed that eq. \ref{eq: pgasf} implies an effective $U_{\rm f}\sim0.01$. For completeness, we repeat their derivation here with our notation. We denote the average energy per ionizing photon as $\hnu$, and the volume density of incident ionizing photons as $\ng\ (= \lion/4\pi r^2 c\hnu)$. From eq. \ref{eq: pgasf} we get
\begin{eqnarray}
 2.3 n_{\rm f}\kB T_{\rm f} &=&  \beta\frac{\lion}{4\pi r^2 c} = \beta \ng \hnu   ~~~~~~ \Rightarrow \nonumber \\
U_{\rm f}\equiv\frac{\ng}{n_{\rm f}} &=& \frac{2.3\kB T_{\rm f}}{\beta\hnu} = 0.03\ T_{f, 4}\ \frac{72\eV}{\beta\hnu}
\label{eq: Uf}
\end{eqnarray}
The numerical value of $\hnu=36\eV$ is appropriate for an ionizing slope of $-1.6$, as seen in luminous AGN (\citealt{Telfer+02}). We emphasize that $\ng$ is measured at the illuminated surface, before any absorption has occurred. 
Using $\ng/n_{\rm f}$ for the effective $U_{\rm f}$ is reasonable, since most of the absorption occurs near the ionization front, where $\tau \sim 1$ and $n\sim n_{\rm f}$ (see D02 and next section). Therefore, eq. \ref{eq: Uf} implies that at $x \sim x_{\rm f}$ RPC gas is similar to constant $n$ gas with an initial ionization parameter $U_0 \equiv \ng/n_0 \sim 0.03$. This effective $U$ is independent of the boundary conditions $n_0$ and $\ng$, and therefore a general property of RPC gas. The value of $U_{\rm f}$ is set only by the ratio of the gas pressure per H-nucleus ($2.3\kB T$) to the pressure per ionizing photon ($\beta \hnu$). 

Using eq. \ref{eq: Uf}, D02 showed that in Seyferts the derived values of $U$ and the small dispersion in $U$ between different objects suggests that the NLR gas is RPC. In Paper II, we use a similar argument to show that the BLR gas is also RPC.

\subsection{The slab structure vs. $\tau$}

\subsubsection{Analytical derivation}

In the optically thin layer at the illuminated surface of the slab, $\frad$ is constant as a function of $x$ and equals to $F_{\rm rad,0}$. Hence, $\frad/c=\prad$, and the hydrostatic equilibrium equation (eq. \ref{eq: hydrostatic only radiation}) can be expressed as 
\begin{equation}\label{eq: hydrostatic optically thin} 
\frac{d \pgas}{dx} = \beta \prad n \sigbar
\end{equation}
Assuming that $\beta$ does not change significantly with $x$, equation \ref{eq: hydrostatic optically thin} can be solved by switching variables to the optical depth $d\tau=n\sigbar d x$:
\begin{equation}
 \frac{d \pgas}{d\tau} = \beta\prad ~~~~~~ \Rightarrow ~~~~~~ \pgas(\tau) = \pgasz + \beta\prad\tau
\end{equation}
For $\pgasz / \beta\prad << \tau << 1$ we get
\begin{equation}\label{eq: pgas vs. prad}
 \pgas(\tau) = \beta\prad\tau
\end{equation}
or equivalently, for $\Xi_0^{-1} << \tau << 1$ we get 
\begin{equation}\label{eq: xi vs tau}
 \Xi(\tau) = \frac{1}{\tau}
\end{equation}
Equation \ref{eq: xi vs tau} implies that in RPC, $\Xi$ has a specific value at each $\tau$, independent of other model parameters. Since the ionization state of the gas is determined to first order by $\Xi$, eq. \ref{eq: xi vs tau} implies a very specific ionization structure for RPC gas, in which the surface layer is highly ionized (high $\Xi$) and ionization decreases with increasing $\tau$. 

For comparison with observations, one needs to know the fraction of the power $W$ emitted in each ionization state. The emission from each layer is equal to the energy absorbed in the layer. However, in a semi-infinite slab only roughly half the power emitted from a certain layer escapes the slab without further absorption. Therefore,
\begin{equation}\label{eq: dtau to dlog xi}
 \frac{ d W(\Xi)}{ d \log \Xi} = 0.5\frac{ d \tau}{ d \log \Xi} = 0.5~\Xi^{-1}
\end{equation}
where the last equality is derived from eq. \ref{eq: xi vs tau}. 
Eq. \ref{eq: dtau to dlog xi} gives directly the fraction of the power emitted in each ionization state, e.g. about $0.5\%$ of the emission comes from $\log\ \Xi\sim 2$ gas, and $\sim5\%$ of the emission comes from $\log \Xi\sim 1$ gas. 

\subsubsection{\cloudy\ calculations}\label{sec: cloudy}

To perform full \cloudy\ calculations, we assume an incident spectrum and gas composition which we consider typical of the AGN and its environment, as follows. We use the \cite{LaorDraine93} SED at $\lambda > 1100$\AA. We assume a power law with index $-1$ at $2-200\kev$ (\citealt{Tueller+08}; \citealt{Molina+09}), and a cutoff at larger frequencies. 
The slope between 1100\AA\ and 2\kev\ is parameterized by $\aeuv$. 
We run models both with and without dust grains, using the the depleted `ISM' abundance set and the default `solar' abundance set, respectively. The actual abundances are scaled linearly with the metallicity parameter $Z$ in all elements except Helium and Nitrogen. For the scaling of the latter two elements with $Z$ we follow G04. We use the dust composition noted as `ISM' in \cloudy, and scale the dust to gas ratio with $Z$. 
We note that \cloudy\ assumes that the radiation pressure on the dust is directly transferred to the gas. 
All \cloudy\ calculations stop at a gas temperature $T \sim 4000\K$, 
beyond which the \hii\ fraction is $<0.1$ and there is only a negligible contribution to the emitted spectrum. 
We use the `constant total pressure' flag, which tells \cloudy\ to increase $\pgas$ between consecutive zones\footnote{\cloudy\ divides the slab into `zones', and solves the local thermal equilibrium and local ionization equilibrium equations in each zone.}, according to the attenuation of the incident continuum (eq. \ref{eq: hydrostatic only radiation}). 

In the dusty models, the calculated pressure due to the trapped line emission is always $<0.03~\pgas$, justifying our assumption in \S\ref{sec: rpc conditions} that it is negligible. However, in the dust-less models the line pressure can be comparable to $\pgas$, which causes stability problems in the \cloudy\ calculation. We therefore turn off the line pressure in the dust-less calculations. In Paper II, we show that including line pressure in the dust-less models changes some of our derived quantities by a factor of $\sim2$. 

In order to compare the \cloudy\ calculations with our analytical derivations above, we need to calculate $x_{\rm f}$, $\tau$, and $\beta$. We set $x_{\rm f}$ to be the $x$ where the H ionized fraction is 50\%. The value of $\tau(x)$ is calculated by summing $\Delta\tau(x')$ on all zones with $x'<x$, where $\Delta\tau(x')=n(x') \sigbar (x')\Delta x'$, and $\Delta x$ is the width of the zone. In dusty models, the dust dominates the opacity, therefore $\sigbar$ is constant to a factor of $\sim2$ at $0<x<x_{\rm f}$. In contrast, in dust-less models $\sigbar$ is dominated by line absorption and bound-free edges, and therefore is a strong function of the ionization state, which changes significantly with increasing $x$ (see below). The value of $\beta$ is derived by comparing the total pressure induced by the radiation at $x_{\rm f}$ to the incident ionizing pressure $\prad$. 

\begin{figure}
\includegraphics{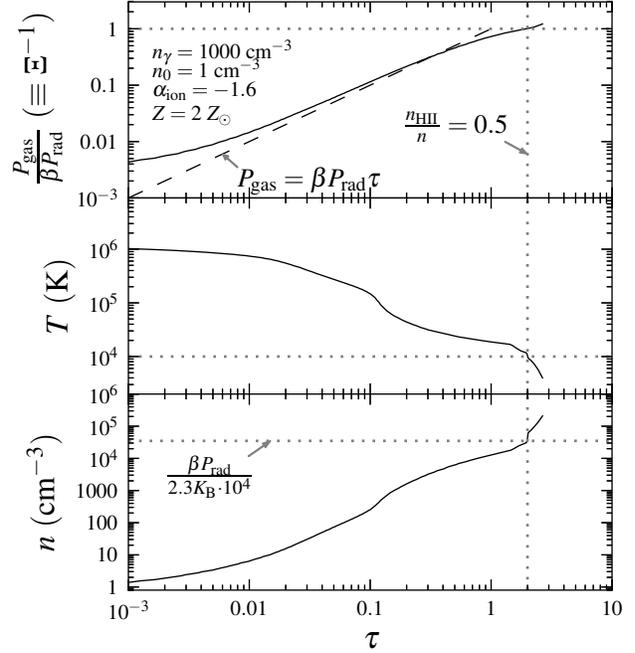}
\caption{
The slab structure of an RPC slab, vs. optical depth. The solid line in each panel shows the result of the \cloudy\ calculation, for the model parameters noted in the top panel. The vertical dotted line marks $\tau_{\rm f}$, where the H ionized fraction is 50\%. 
{\bf (Top)} The $\pgas$ structure. 
With the absorption of ionizing radiation, $\pgas$ increases from $\pgasz$ to $\pgasf=\beta\prad$ (eq. \ref{eq: pgasf}, horizontal dotted line).  
In this dusty model $\beta=1.9$ due to the absorption of optical photons by dust grains. The dashed line marks eq. \ref{eq: pgas vs. prad}, which is the analytical derivation of the slab structure assuming $\pgasz\rightarrow0$ and $\tau<<1$. Eq. \ref{eq: pgas vs. prad} is similar to the \cloudy\ calculation at $0.01\lesssim\tau\lesssim1$. 
{\bf (Middle)} The $T$ structure, which drops from $10^6\K$ at $\tau=0$ to $10^4\K$ (horizontal dotted line) at $\tau=\tau_{\rm f}$. Over 80\% of the absorption occurs at $T<40\,000\K$. 
{\bf (Bottom)} The structure of $n$, which increases by four orders of magnitude at $0<\tau<1$, due to the increase in $\pgas$ and the drop in $T$. The large range in $n$ implies a large range in $U$ within a single slab. The horizontal dotted line marks the expected $n_{\rm f}$ (eq. \ref{eq: n vs r}).
}
\label{fig: profile vs tau}
\end{figure}

Figure \ref{fig: profile vs tau} shows the slab structure of a dusty model, with $\aeuv=-1.6$, typical of luminous AGN (\citealt{Telfer+02}), and $Z=2\zsun$, the $Z$ observed in the ISM of quiescent galaxies with stellar masses $\sim 10^{11}\msun$, which is likely the $Z$ also found in the NLR of a typical AGN host (\citealt{Groves+06}; \citealt{SternLaor13}). We use $\ng=1000\cm^{-3}$, which corresponds to $r=70~L_{{\rm i,} 45}^{1/2}\pc$. Different values of $\aeuv,\ Z$ or $\ng$ do not affect the conclusions of this section. We assume $n_0=1\cm^{-3}$, so $U_0=1000$ and $\beta\prad/\pgas=\Xi=300$, well within the RPC regime (eq. \ref{eq: prad>pgas}). 
The $\beta$ in this model is equal to $1.9$.

The top panel shows that $\pgas$ increases from the assumed $\pgasz$ at $\tau=0$ to $\beta\prad$ at $\tau=\tau_{\rm f}$ (eq. \ref{eq: pgasf}). 
Equation \ref{eq: pgas vs. prad} (dashed line), which is the analytical derivation of the slab structure assuming $\pgasz\rightarrow0$ and $\tau<<1$, is a good approximation of the \cloudy\ calculation at $0.01\lesssim\tau\lesssim1$.  
Equivalently, $\Xi$ decreases from $300$ at $\tau=0$ to $1$ at $\tau=\tau_{\rm f}$, as expected from eqs. \ref{eq: xi vs tau} and \ref{eq: Xi_f}. 
Therefore, the analytical derivations of the slab structure vs. $\tau$ agree with the full \cloudy\ calculations. 

The middle panel of Fig. \ref{fig: profile vs tau} shows the $T(\tau)$ profile, which drops from $10^6\K$ at $\tau=0$, to $10^4\K$ (dotted line) at the ionization front. Note that $T=40\,000\K$ at $\tau=0.2$, therefore 80\% of the absorption occurs in gas with $T<40\,000\K$. The bottom panel shows that $n$ increases by four orders of magnitude at $0<\tau<1$, reaching the expected $n_{\rm f}$ (eq. \ref{eq: n vs r}) at the ionization front. This large increase in $n$ is due to the increase in $\pgas$ and the drop in $T$.  The large change in $n$ results in a large drop in $U$, from $U_0=1000$ to $U_{\rm f}\approx0.03$ (eq. \ref{eq: Uf}). This large range in $U$ is apparent in the Oxygen and Neon ionization structure within the slab, presented in Appendix \ref{app: ionization structure}. 

We note that the exact $T(\tau)$ and $n(\tau)$ profiles depend on the details of the dust physics and its interaction with the gas, which are subject to some uncertainty. Specifically, the solution at $T\sim10^5\K$ may not be unique, and there could be two phases at the same pressure. We ignore this possibility here. However, the increase of $\pgas$ with $\tau$, which is the main conclusion of this section, is independent of the exact $T(\tau)$ profile.

Eq. \ref{eq: pgas vs. prad} suggests that $\pgas(\tau)$ is independent of the source of opacity, and therefore the $\pgas(\tau)$ profile of dusty and dust-less models should be similar. Indeed, we find that the $\pgas(\tau)$ of dust-less models are similar to the $\pgas(\tau)$ of dusty models seen in the top panel of Fig. \ref{fig: profile vs tau}. We address the effect of dust in a more detailed manner in the following section, where we analyze the slab structure as a function of $x$, where the effect of dust is more prominent.

\subsection{The slab structure vs. $x$}

\subsubsection{The pressure scale length}\label{sec: lp}

Eq. \ref{eq: hydrostatic optically thin}, which assumes $\tau<<1$, can be rewritten as 
\begin{equation}\label{eq: pgas differential}
\frac{d \pgas}{dx} = \beta\prad \sigbar\frac{\pgas}{2.3\kB T} 
\end{equation}
In order to tract the problem analytically, we assume that $\sigbar$ and $T$ do not change with $x$. The accuracy of this approximation will become apparent below, where we compare the analytical result to the full \cloudy\ calculation.
Hence, eq. \ref{eq: pgas differential} can be integrated to 
\begin{equation}
\pgas(x) = \pgasz ~ e^{x/\lp} 
\label{eq: exponential P}
\end{equation}
or equivalently, 
\begin{equation}
 \Xi(x) = \Xi_0 ~ e^{-x/\lp}
\end{equation}
The {\it pressure scale length}, $\lp$, is equal to 
\begin{equation}
\lp = \frac{2.3\kB T}{\beta\prad\sigbar} = 0.9 ~\sigbar_{-21}^{-1}T_6 \frac{r_{50}^2}{\beta L_{{\rm i,} 45}}  \pc 
\label{eq: lp1}
\end{equation}
where $T=10^6T_6\K$, the appropriate $T$ at the illuminated surface (Fig. \ref{fig: profile vs tau}), and $\sigbar=10^{-21}\sigbar_{-21}\cm^2$. We show below that this value of $\sigbar$ is typical of dusty gas, but is significantly lower in dust-less gas.
Equivalently, $\lp$ can be expressed with $\ng$ and $\hnu$:
\begin{equation}
\lp = \frac{2.3 \kB T}{\beta\hnu\ng\sigbar} = 1.8 ~\sigbar_{-21}^{-1} T_6 \frac{72\eV}{\beta\hnu} n_{\gamma, 3}^{-1} \pc 
\label{eq: lp2}
\end{equation}
where $\ng=1000~n_{\gamma, 3}\cm^{-3}$.

Within the slab, $U$ is lower than at the illuminated surface. 
The decrease in $U$ implies that $T$ can only be lower within the slab than at the surface, and we show below that $\sigbar$ can only be higher within the slab than at the surface. Therefore, eqs. \ref{eq: lp1}--\ref{eq: lp2} imply that the largest $\lp$ in the slab is at the illuminated surface. 

\begin{figure*}
\includegraphics{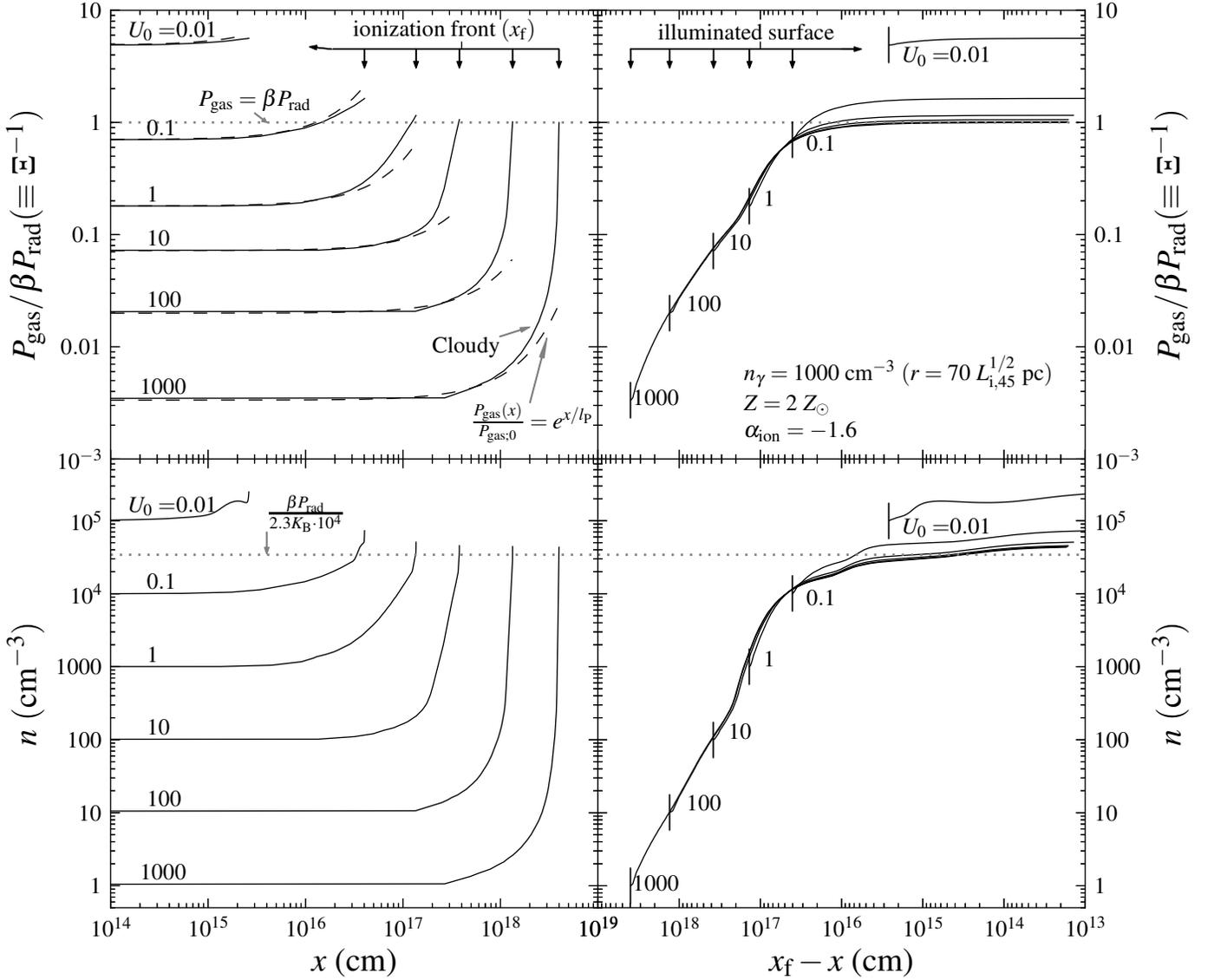}
\caption{
The structure of a dusty RPC slab vs. distance from the illuminated surface (left) and vs. distance from the ionization front (right).
The solid lines show the results of the \cloudy\ calculations at $0<x<x_{\rm f}$, for models with different $U_0$ (noted near each line). 
Other model parameters are noted in the top-right panel. 
{\bf (Left panels)} 
The value of $\pgas$ and $n$ increase with $x$ as radiation is absorbed by the gas. 
Models with $U_0>>0.03$ (or $\Xi_0>>1$) have $\prad>>\pgasz$, and are therefore RPC. 
The $\pgasf$ in all RPC models is independent of $U_0$, and equal to $\beta\prad$ (dotted line in the top panels). 
Similarly, in all RPC models $n_{\rm f}$ is independent of $U_0$ and given by eq. \ref{eq: n vs r} (dotted line in the bottom panels). 
In the $U_0=0.01$ model $\pgasz>\prad$, therefore this model is not RPC and requires an additional source of confinement. 
Dashed lines in the top-left panel plot eq. \ref{eq: exponential P}, the analytical expression for $\pgas(x)$. The increase in $\pgas$ becomes significant at $x\sim\lp$ (eq. \ref{eq: lp1}), where $\lp$ depends on $U_0$. 
The analytical and \cloudy\ calculations give similar results. 
{\bf (Right panels)} 
The illuminated surface of each model is noted by a `$\mid$'. All RPC models lie on the same $\pgas(x_{\rm f}-x)$ and $n(x_{\rm f}-x)$ profiles. That is, two RPC models with different boundary conditions $n_0=n'$ and $n_0=n''>n'$ have the same solution at $n>n''$, and differ only in the existence of an optically thin surface layer with $n'<n<n''$. Therefore, RPC solutions are essentially independent of the boundary value $n_0$ or $U_0$.
}
\label{fig: different P0}
\end{figure*}

\begin{figure*}
\includegraphics{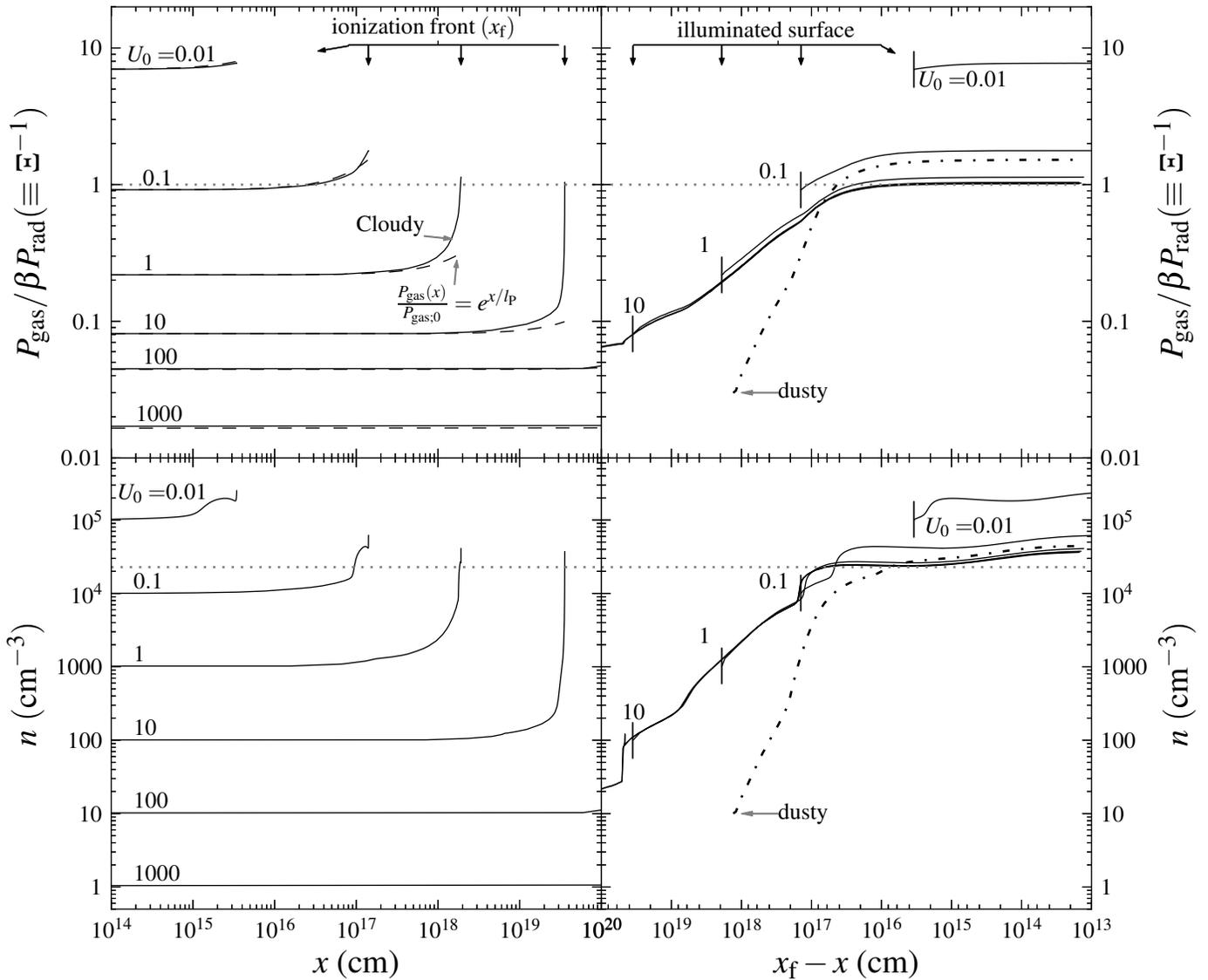}
\caption{
Same as Fig. \ref{fig: different P0}, for dust-less models. 
{\bf (Left panels)} 
Models with $0.03 << U_0 \leq 10$ are RPC. With absorption of radiation, $\pgas$ and $n$ increase, reaching $\pgasf=\beta\prad$ and $n_{\rm f}\approx\beta\prad/2\kB\cdot 10^4$ (dotted horizontal lines), with $\beta=1.2$. 
The $\lp$ increases with $U_0$, due to the associated increase in $T_0$ and the decrease in $\sigbar_0$ (eq. \ref{eq: lp1}). 
In models with $U_0 \geq 100$, $\lp > r \sim 70\pc$, and there is no significant increase in $\pgas$ at $x<r$. 
{\bf (Right panels)} 
As in the dusty models (Fig. \ref{fig: different P0}), all dust-less models with $U_0>>0.03$ lie on the same $\pgas(x_{\rm f}-x)$ and $n(x_{\rm f}-x)$ profile.
The dust-less models and the dusty model (dash-dotted lines) have a similar $\pgasf$, differing only due to their different $\beta$. The increase in $n$ and $\pgas$ in the surface layer of dust-less models occurs on scales $100$ times larger than in the dusty models, mainly due to the lower $\sigbar$. 
}
\label{fig: different P0 no dust}
\end{figure*}

The \cloudy\ calculations of the slab structure at $0<x<x_{\rm f}$ are shown in the left panels of Figures \ref{fig: different P0} and \ref{fig: different P0 no dust}. As in Fig. \ref{fig: profile vs tau},  we use $\aeuv = -1.6$, $Z=2\zsun$, and $n_{\gamma, 3}=1$. The conclusions of this section are robust to other reasonable choices of $\aeuv$ and $Z$, while the effect of changing $\ng$ is addressed in the following section. We analyze models with different $n_0$ (or equivalently different $U_0$ or different $\pgasz$). 

Fig. \ref{fig: different P0} shows the calculation of the dusty models. In the left panels, all models with $U_0>>0.03$ show a similar behavior. From a certain scale which is different in each model, $\pgas$ and $n$ significantly increase, reaching the same $\pgasf=\beta\prad$ (eq. \ref{eq: pgasf}) and $n_{\rm f}\approx\beta\prad/2\kB\cdot 10^4$ (eq. \ref{eq: n vs r}) at $x=x_{\rm f}$. Therefore, the conditions at $x=x_{\rm f}$ are independent of the conditions at $x=0$.

For comparison, we also calculate the analytical expression for $\pgas(x)$ (eq. \ref{eq: exponential P}), which requires an evaluation of $\lp$ at the illuminated surface. The value of $\lp$ depends on $\sigbar$, which is dominated by the dust opacity at $U_0>>0.006$ (\citealt{NetzerLaor93}), and equals to
\begin{equation}\label{eq: dusty sigbar}
\sigbar_{\rm dust}=0.8\times 10^{-21} \frac{Z}{\zsun}\cm^{-2}
\end{equation}
for the assumed $\aeuv=-1.6$. We emphasize that this $\sigbar$ is independent of $U_0$ for $U_0>>0.006$. 
Hence, for the models in Fig. \ref{fig: different P0}, using eq. \ref{eq: lp2} we get $\lp = 1.8\times10^{18}~T_6 \cm$. 
The dashed lines in the top-left panel of Fig. \ref{fig: different P0} plot the analytical $\pgas(x)$. The analytical and \cloudy\ calculations of $\pgas(x)$ agree rather well. With increasing $U_0$, $\lp$ increases due to the increase in $T_0$. At $x>\lp$, the analytical expression somewhat underestimates $\pgas(x)$, due to the decrease in $T$, which is not accounted for in the analytical derivation. 

In the $U_0=0.01$ model $\pgasz>\prad$, therefore $\pgasf \approx \pgasz$. This model is not RPC, since an additional source of confinement which is stronger than the radiation pressure is required to achieve $\pgas>\prad$. The small dynamical range of $n$ in this model implies that it is effectively a constant-$n$ model. 

Fig. \ref{fig: different P0 no dust} shows the results of the \cloudy\ calculation of the dust-less models. 
The models with $U_0 \leq 10$ behave in a similar fashion as the dusty models. At $x\approx\lp$, $\pgas(x)$ significantly increases, reaching $\pgasf=\beta\prad$ at $x=x_{\rm f}$. Here, $\beta=1.2$. 
The $\lp$ in the dust-less models are larger than in the dusty models, mainly due to the lower $\sigbar$
\footnote{Also, $T_0$ is lower in dusty models due to the cooling provided by dust-gas interactions. The effect of the different $T$ on $\lp$ is small compared to the effect of the different $\sigbar$.}. 
We find that at $U > 100$
\begin{equation}\label{eq: dust-less sigbar}
\sigbar_{\rm es}=\frac{n_e}{n}\sigma_{\rm Th} = 0.9\times10^{-24}\cm^2
\end{equation}
where $\sigma_{\rm Th}$ is the Thompson cross section. For lower $U$, line and edge opacity surpasses the electron scattering opacity, and $\sigbar$ increases. In the $U_0=1000$ model, the total column density at $0<x<x_{\rm f}$ is $3.6\times10^{22}\cm^{-2}$, implying that $\sigbar$ averaged over the ionized layer is $10^{-22.5}\cm^2$, a factor of 60 lower than the value of $\sigbar$ in dusty gas (eq. \ref{eq: dusty sigbar}). 

In models with $U_0 \geq 100$, $\lp > 70\pc$, which is larger than $r$ for the assumed $\ng$ in an AGN with $L_{{\rm i,} 45}=1$. Therefore, our assumption that $x<<r$ is violated. We address this constraint in \S\ref{sec: ng}.

\subsubsection{The slab structure vs. $x_{\rm f} - x$}

Above we showed that the $\pgas(x)$ profiles of RPC models with different $n_0$ differ, since $\lp$ increases with $U_0\propto n_0^{-1}$. 
However, Fig. \ref{fig: profile vs tau} shows that $n$ increases significantly already at $\tau<<1$. Therefore, if we compare two RPC models with different boundary conditions $n_0=n'$ and $n_0=n''>n'$, the former model should have an optically thin surface layer in which $n'<n<n''$. Since this surface layer is optically thin, we do not expect its existence to significantly affect the solution at $n>n''$. In the inner layer where $n>n''$, the two RPC models should have similar solutions. 

A similar solution at $n>n''$ implies that if we present the slab structure as a function of distance from some depth {\it within} the slab, such as $x_{\rm f} - x$, then solutions of models with different $n_0$ should be practically identical, differing only in their starting point. In the right panels of Fig. \ref{fig: different P0} we show the slab structure of the dusty models vs. $x_{\rm f} - x$. 
The illuminated surface of each model is noted by a `$\mid$'. All RPC models ($U_0>>0.03$) lie on the same $\pgas(x_{\rm f}-x)$ and $n(x_{\rm f}-x)$ profiles, differing only in the $x_{\rm f}-x$ value of the illuminated surface, where models with lower $n_0$ extend to larger distances from the ionized front. Therefore, RPC models with different $U_0$ have a very similar slab structure. Since most of the line emission comes from parts of the slab with $n \lesssim n_{\rm f}$ (see below), which Fig. \ref{fig: different P0} shows is common to all RPC solutions, it follows that RPC solutions are essentially independent of the boundary value $n_0$ or $U_0$.

The right panels of Fig. \ref{fig: different P0 no dust} show that the dust-less models all lie on the same $\pgas(x_{\rm f}-x)$ and $n(x_{\rm f}-x)$ solution, as seen in the dusty models in Fig. \ref{fig: different P0}. Therefore, our conclusion that the slab structure is insensitive to $n_0$ is independent of the dust content of the gas. The effect of dust is apparent in two main aspects. The physical length of the optically thin surface layer is smaller by a factor of $\sim100$ in the dusty models, mainly due to the increase in $\sigbar$. Also, $\beta$ and $\pgasf$ are larger by 50\% in the dusty models, due to the additional pressure from absorption of optical photons by the dust.

\subsection{The slab structure vs. $\ng$}\label{sec: ng}

\begin{figure*}
\includegraphics{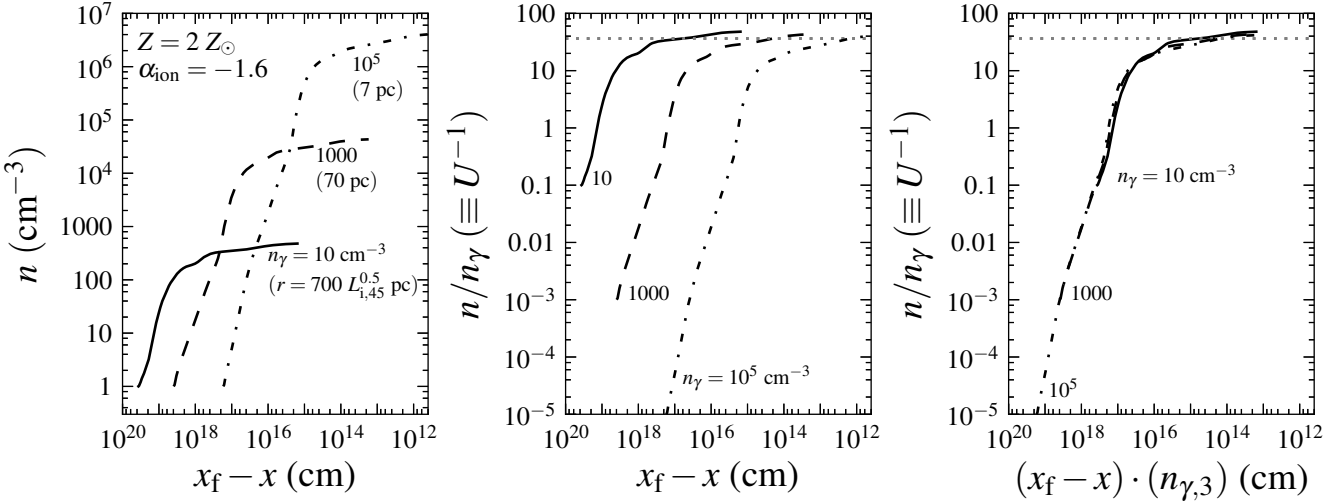}
\caption{The density structure of RPC slabs with different $\ng$. 
Each line denotes an RPC dusty model with a different $\ng$ (noted), or equivalently a different $r$ (noted in left panel). Other model parameters are noted in the left panel. 
{\bf (Left)} The $n(x_{\rm f}-x)$ profile. The $n$ increases from the assumed $n_0=1\cm^{-3}$ to $n_{\rm f}\propto n_\gamma$. 
{\bf (Middle)} The ordinate is normalized by $\ng$. All models reach the same $U_{\rm f}$ (eq. \ref{eq: Uf}, dotted line).
{\bf (Right)} The ordinate is normalized by $\ng$ and the abscissa is normalized by $(\ng/1000\cm^{-3})^{-1}$. 
Models with different $\ng$ lie on the same $n/\ng$ vs. $(x_{\rm f}-x)\ng$ profile. Therefore, the slab structure of RPC gas with different $\ng$ is similar if $n$ is scaled by $\ng$ and $x$ is scaled by $\ng^{-1}$. 
}
\label{fig: different ngamma}
\end{figure*}

For the assumed SED,
\begin{equation}\label{eq: ng}
\ng = 2040~\frac{L_{\rm i,45}}{r_{50}^2}\frac{36\eV}{\hnu}\cm^{-3} 
\end{equation}
How does the slab structure depend on $\ng$? or equivalently, for a given $\lion$, how does the slab structure depend on $r$ ?

At low $n$, the ionization state and $T$ of the gas are a function mainly of $U$, while the direct dependence on $n$ and $\ng$ are only a second-order effect. This insensitivity of the ionization state to $n$ and $\ng$ follows from the fact that the ionization rate is $\propto \ng n$, while the recombination rate is $\propto n^2$. Therefore the ionization balance is mainly a function of $\ng/n \equiv U$. 
Similarly, the heating rate depends on $\ng \hnu n$, while collisional cooling is $\propto n^2$. Hence, for a given SED the $T$ balance is also determined to first order by $U$\footnote{A notable exception is Compton cooling, which is $\propto \ng \hnu n$. In this case the $T$ balance is independent of either $n$, $\ng$ or $U$, and determined solely by the SED.}.

The above reasoning assumes a fixed dust content in the gas, since a changing dust content with $\ng$ will create a direct relation between the slab structure and $\ng$. This assumption is violated at $10^{5.9}<\ng<10^{8.6}\cm^{-3}$, where different dust grain species sublimate at different $\ng$ (\S\ref{sec: dust destruction}). 

To understand the effect of $\ng$ on the slab structure, we examine its effect on $\lp$, which determines the physical scale of the slab. Eq. \ref{eq: lp2} shows that the value of $\lp \cdot \ng$ depends on $\beta, T, \hnu$ and $\sigbar$. The reasoning above implies that $T$ is a function of $U$. The values of $\sigbar$ and $\beta$ depend on the ionization state,  $T$, and dust content of the gas, and are therefore also mainly a function of $U$ at $\ng<10^{5.9}\cm^{-3}$. Hence, for a given $\hnu$ we can write
\begin{equation}\label{eq: different ng}
\lp\ng = \mathcal{F}(U) = \mathcal{F}(\frac{\ng}{n})
\end{equation}
where $\mathcal{F}(U)$ is a computable function of $U$. 
Eq. \ref{eq: different ng} suggests that solutions of models with different $\ng$ are equivalent, if we scale $n$ by $\ng$, and scale $x$ by $\ng^{-1}$.

In Figure \ref{fig: different ngamma}, we show the validity of eq. \ref{eq: different ng} using \cloudy\ calculations. The left panel shows $n$ vs. $x_{\rm f} - x$ of dusty RPC slabs with different values of $\ng$. We assume $n_0=1 \cm^{-3}$ in all models. The value of $n$ increases as radiation pressure is absorbed, reaching a $n_{\rm f}$ which increases with increasing $\ng$. 
In the middle panel $n$ is scaled by $\ng$. All models reach the same $n_{\rm f}/\ng$, as expected from eq. \ref{eq: Uf}. 
In the right panel we also scaled $(x_{\rm f}-x)$ by $\ng^{-1}$. The $U$ vs. $(x_{\rm f}-x)\ng$ profiles of the different models are almost identical, as implied by eq. \ref{eq: different ng}. The models differ only in the starting point of their solution, due to the different assumed $U_0$. Therefore, the slab structure of dusty RPC models with different $\ng$ are similar once $x$ is scaled by $\ng^{-1}$ and $n$ is scaled by $\ng$. Dust-less RPC models with different $\ng$ show the same property, as do RPC models of warm absorbers (\citealt{Chevallier+07}).

We can now derive the range in $r$ where the slab approximation ($x_f<<r$) is valid. 
Since $\pgas$ increases exponentially with a scale of $\lp$ (eq. \ref{eq: exponential P}), and since the largest $\lp$ is at the surface, then $x_{\rm f}\lesssim\ln(n_{\rm f}/n_0)\cdot l_{\rm p,0}$. 
We assume $U_0=100$, as material at higher $U$ does not emit emission lines, so from eq. \ref{eq: Uf} we get $\ln(n_{\rm f}/n_0) = \ln(U_0/U_{\rm f}) = 8$. 
In the dusty $\ng=1000\cm^{-3}$ model we find $\sigbar_{-21}=1.6, T_6=0.7$ and $\beta=1.9$ at the surface. The above discussion suggests that these properties are independent of $r$. Therefore, by plugging these values in eq. \ref{eq: lp1} we get
\begin{equation}
\frac{x_{\rm f,dusty}}{r} \lesssim\frac{8~\l_{\rm p,0}}{r} =  0.03~L_{\rm i,45}^{-1}r_{50}
\end{equation}
implying that in dusty RPC gas the slab approximation is valid at least up to $\kpc$ scale. 
In contrast, at the surface of the $U_0=100$ dust-less model we find $\sigbar = \sigbar_{\rm es}, T_6=1$ and $\beta=2.9$, so 
\begin{equation}
\frac{x_{\rm f, dust{\text -}less}}{r} \lesssim \frac{8~\l_{\rm p,0}}{r} =  55~L_{\rm i,45}^{-1}r_{50}
\end{equation}
implying that the slab approximation is invalid for dust-less gas on NLR scales, if $U_0=100$. The right panels of Fig. \ref{fig: different P0 no dust} show that at $r=70~L_{\rm i,45}\pc$, the slab approximation will be valid only if $U_0<10$.

\subsection{Emission line emissivity vs. $r$}\label{sec: emissivity vs. r}
\newcommand{\lline}{L_{\rm line}}

\begin{figure*}
\includegraphics{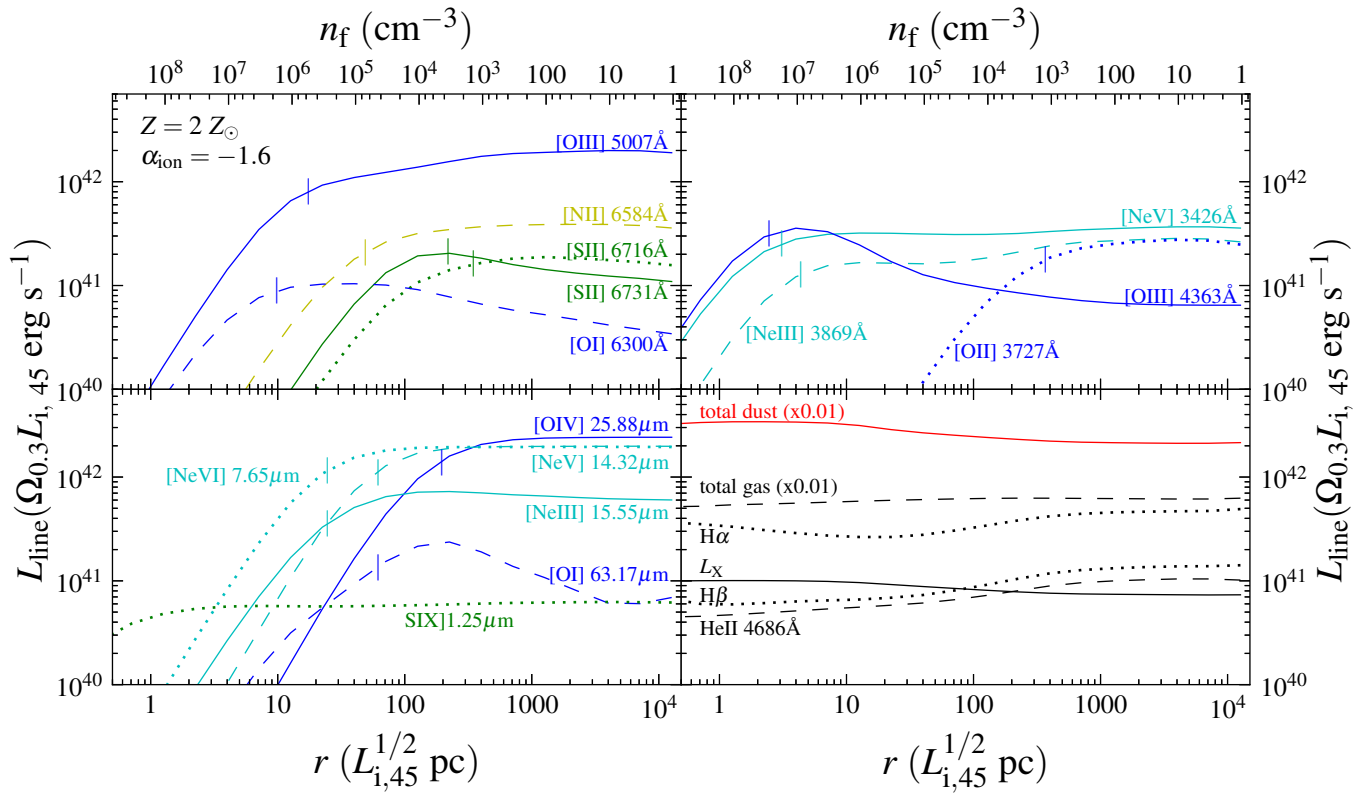}
\caption{The emission line luminosity of RPC slabs, vs. distance from the nucleus.
Plotted lines denote $\lline$ at different $r$, for $L_{{\rm i,} 45}=1$ and a covering factor $\Omega(r)=0.3$. The $\lline$ are calculated by \cloudy\ using dusty gas and the model parameters noted in the top-left panel. The line marked `$\lx$' includes all line emission at $0.5 - 2 \kev$. Also shown are the total emission from the dust grains and the total emission from the gas, scaled down by a factor of 100 (lower right panel). The $n_f$ (noted on top) is set by $r$ via eq. \ref{eq: n vs r}. 
The approximate EW of an optical line with wavelength $\lambda$ can be derived by dividing the y-axis value by $6.1\times10^{40} (\lambda/5007{\rm \AA})^{-1.5} \ergs \AA^{-1}$. 
About 80\% of the emission is thermal IR emission from dust grains, because dust dominates the opacity in RPC gas. 
A wide range of gas ionization states is observed at each $r$, from \oi-emitting layers to \six\ and soft X-ray emitting layers. 
The $\lline$ of recombination lines, and the total gas and dust luminosities, are constant with $r$ up to a factor of $\sim2$, due to the similar slab structure at different $r$ (Fig. \ref{fig: different ngamma}). 
The $\ncrit$ of forbidden lines are marked by `$|$'. At $n_f>>\ncrit$, forbidden lines exhibit $\lline\propto n_f^{\sim-1}\propto r^{\sim 2}$ due to collisional de-excitation. At $n_f<<\ncrit$, forbidden lines exhibit radial dependencies between $\lline\propto r^{-0.6}$ and $\lline\propto r^0$. 
}
\label{fig: ew vs r}
\end{figure*}

In the previous sections, we showed that RPC slabs have $n\propto r^{-2}$. 
In this section, we use this result to predict the emission line luminosities $\lline$ as a function of $r$. 

To calculate $\lline$, we use \cloudy, as described in \S\ref{sec: cloudy}. We run a grid of dusty \cloudy\ models with $\ng=10^{-1.5}-10^{8}\cm^{-3}$, which corresponds to $r=0.2-10^4 ~ L_{\rm i,45}^{1/2}\pc$ (eq. \ref{eq: ng}). We set $U_0=10^4$, to reside deep in the RPC regime. Identical results are found for all $U_0\geq 1$, as implied by Fig. \ref{fig: different P0}. We set $Z=2\zsun$ and $\aeuv=-1.6$ (see \S\ref{sec: cloudy}). Our choice of a dusty model is motivated by the observations that the emission line gas is likely dusty at $\ng<10^{8.5}\cm^{-3}$, at least in layers which emit emission lines with IP$\sim40\eV$ or less (\S\ref{sec: dust destruction}). 
The plotted $\lline$ of lines with IP $>>40\eV$ may be inaccurate if dust is significantly destroyed in the layers in which they are emitted. Also, we disregard the change in dust composition due to sublimation of small grains at $10^{5.9}<\ng<10^{8.5}\cm^{-3}$
(see \S\ref{sec: dust sublimation}). Since we assume the slab is semi-infinite, all emission properties are measured at the illuminated surface. The back side of the slab should mainly emit dust thermal IR emission. 

The value of $\lline$ of different emission lines for different $r$ are shown in Figure \ref{fig: ew vs r}. We mark the sum of emission of lines with energies $0.5 - 2 \kev$ by $\lx$\footnote{The $\lx$ is dominated by recombination and resonance lines, as expected for photoionized gas.} (lower right panel). Also shown are the total emission from the dust grains and the total emission from the gas. 
We assume $L_{{\rm i,} 45}=1$, and a covering factor at $r$, $\Omega(r)$, of $0.3$. The $\lline(r)$ for other values of $L_{{\rm i,} 45}$ and $\Omega(r)$ can be derived with the appropriate scaling. The approximate equivalent width (EW) of an optical line with wavelength $\lambda$ can be derived by dividing the y-axis value by $6.1\times10^{40} (\lambda/5007{\rm \AA})^{-1.5} \ergs \AA^{-1}$. 

Several properties of the emitted spectrum can be deduced from Fig. \ref{fig: ew vs r}.
Emission lines from a wide range of ionization states are apparent, from the coronal and soft X-ray lines emitted from the surface layer, to the \sii\ and \oi\ lines emitted from the partially ionized layer. 
The luminosities of all shown recombination lines, and the total gas and dust luminosities (lower right panel), are constant with $r$ up to a factor of $\sim2$. 
This similarity is due to the similar slab structure at different $r$ (Fig. \ref{fig: different ngamma}), and because these emission properties are independent of  $n$. The fraction of emission in dust IR thermal emission is 77-87\%. This high fraction is because in RPC, most of the absorption occurs at $U> U_f=0.03$ (eq. \ref{eq: Uf}), where dust dominates the opacity. 
The dominant trend with decreasing $r$ is the collisional de-excitation of forbidden lines. At $n_{\rm f}>>\ncrit$, $\lline \propto \sim n_{\rm f}^{-1} \propto \sim r^{2}$.  The luminosities of $\oiii\ 4363\AA$, $\oi\ 6300\AA$ and $\oi\ 63.17\mic$ peak at $n_{\rm f}\sim \ncrit$, while the luminosities of most other forbidden lines actually remains constant at $n_f<<\ncrit$.

\begin{figure*}
\includegraphics{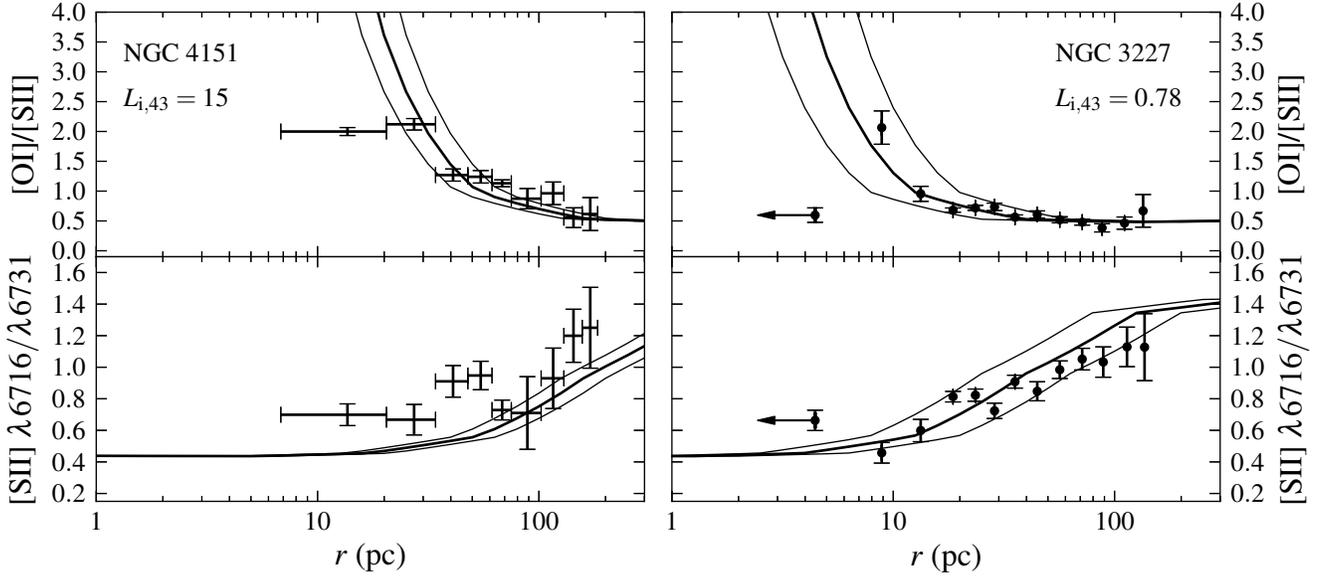}
\caption{The expected and observed NLR gas density as a function of distance.
The error bars are \HST\ observations of the $n_{\rm f}$-sensitive $\oi/\sii$ and $\sii\ 6716/6731$ ratios from Kraemer et al. (2000a, NGC~4151, left panels) and from Walsh et al. (2008, NGC~3227, right panels). The center measurement in NGC~3227 is marked by a left-pointing arrow. 
The black solid line in each panel plots the expected line ratios for RPC slabs at different $r$. The calculations are performed with \cloudy, with no free parameters. The gray lines denote the uncertainty due to the assumed $0.2$ dex uncertainty in the $\lion$ estimate, which is noted in the top panels. 
Except the central measurement, the observed $\oi/\sii$ ratios agree with the RPC calculations.
In NGC~4151, the expected $\lambda6716/\lambda6731$ are typically lower than observed. In NGC~3227, the observed $\lambda6716/\lambda6731$ imply a somewhat flatter $n_{\rm f}$ vs. $r$ relation than expected by RPC. We suspect that since $\lambda6716/\lambda6731$ is sensitive only to a small dynamical range in $n$, projection effects in the observations may hinder the ability of $\lambda6716/\lambda6731$ to estimate the true $n_{\rm f}$ at each $r$. 
}
\label{fig: SII vs r}
\end{figure*}

Weaker trends include the decrease in \Hb\ and \Heii\ 4686\AA\ by a factor of two with decreasing $r$. This decrease is because $T$ increases at lower $r$, due to the collisional suppression of the main coolants. The increase in $T$ induces a higher $U$ (see eq. \ref{eq: Uf}), which increases the ratio of dust to gas opacity, thus decreasing the amount of ionizing photons absorbed by H and He. This trend disappears in the dust-less models, where the emissivity of recombination lines remains constant with $r$. However, the model ignores sublimation of the smaller grains at small $r$, which will decrease the dust to gas opacity compared to large $r$, and will affect the line strength. 

We note that RPC implies that one cannot define a boundary between the so-called `torus' and the NLR. The lower right panel of Fig. \ref{fig: ew vs r} shows that the dust IR thermal emission per unit $\Omega$ is nearly the same on `torus scales' (immediately beyond the sublimation radius) and on `NLR scales' ($10\pc-1\kpc$). Both thermal IR dust emission and recombination lines are emitted at all $r$, with a nearly constant emissivity per unit $\Omega$. Only specific forbidden lines cannot be emitted at small enough $r$, 
depending on their $\ncrit$, due to collisional de-excitation.

Below, we compare the results of Fig. \ref{fig: ew vs r} with both resolved and unresolved observations of the NLR. In order to compare RPC with unresolved observations, we need to know $\Omega(r)$, in order to derive the value of $\lline$ integrated over all $r$. For simplicity, we parameterize $\Omega(r)$ as a power-law:
\begin{equation}\label{eq: eta}
 \frac{ d \Omega(r)}{ d \log r} \propto r^\eta ~~~ (r_{\rm in} < r < r_{\rm out})
\end{equation}
Now, the emissivity of forbidden lines scales as $r^2$ at $n>>\ncrit$, and is constant or decreasing with $r$ at $n<<\ncrit$ (Fig. \ref{fig: ew vs r}). Therefore, eq. \ref{eq: eta} implies that if $-2 < \eta < 0$, then the integrated $\lline$ will be dominated by emission from $r$ such that $n_{\rm f}(r)\sim\ncrit$.

A constraint on $\eta$ can be derived from the flat IR $\nu L_\nu$ slope observed in quasars at $10^{13}<\nu<10^{14}\Hz$ (e.g. \citealt{Richards+06}). This flat slope suggests that the dust thermal emission per unit $\log\ T_{\rm dust}$ is constant at $100K<T_{\rm dust}<1000K$. To first order, $T_{\rm dust}$ is proportional to the effective temperature of the radiation (see fig. 8 in \citealt{LaorDraine93}), which is $\propto r^{-1/2}$. The emission at each $r$ is $\propto d\Omega(r)$. Therefore, the flat IR slope implies $ d \Omega(r)/ d(\log r^{-1/2})\propto \sim r^{0}$. Using the definition of $\eta$, this relation implies that $\eta\sim0$, at $r$ where $T_{\rm dust}(r)>100\K$.

\section{Comparison with observations}\label{sec: observations}

RPC provides robust predictions on the gas properties as a function of distance in AGN. Are these properties observed?

\subsection{Resolved observations of the $n_{\rm f}$ vs. $r$ relation}\label{sec: resolved}

We compare eq. \ref{eq: n vs r} with \HST\ observations of two Seyferts in the literature, NGC~4151 which was observed by \cite{Kraemer+00}, and NGC~3227 which was observed by \cite{Walsh+08}. To estimate $\lion$, we use the \cite{Laor03} bolometric luminosity ($\lbol$) estimates, which are based on the \cite{HoPeng01} B-band nuclear magnitude measurements taken by \HST. 
In order to estimate the observed $n_{\rm f}$, we use the $\oi\ \lambda6300/\sii\ \lambda6716$ ratio (e.g. \citealt{Barth+01}). These two low ionization lines are both emitted from the partially ionized region of the slab, where $n\sim n_{\rm f}$. However, because these lines differ significantly in $\ncrit$, their luminosity ratio is sensitive to $n$ at $10^{2.5} \lesssim n \lesssim 10^{6.5}\cm^{-3}$ (Fig. \ref{fig: ew vs r}). 
We use models with $Z=2\zsun$ and $\aeuv=-1.4$, typical of AGN with Seyfert luminosities (\citealt{Steffen+06}). 
We set $\lion=0.35~\lbol$, appropriate for our assumed SED. We find $L_{\rm i,43}=15$ and $L_{\rm i,43}=0.78$ for NGC~4151 and NGC~3227, respectively, where $\lion=10^{43}L_{\rm i,43}\ergs$. We assume an error of $0.2$ dex in the $\lion$ estimate. 
\cite{Walsh+08} also observed \oi\ in four LINERs (\citealt{Heckman80}), though the incident SED and $\lion$ are not well-constrained in LINERs, so a quantitative comparison with RPC is less reliable. We discuss LINERs in the context of RPC in \S \ref{sec: implications}. 

The expected and observed $\oi/\sii$ are compared in the top panels of Figure \ref{fig: SII vs r}. In NGC~4151, we average the observations of the South-West and North-East sides of the slit. In NGC~3227, we average over all position angles in bins of 0.1 dex in $r$. We note that the central measurement in NGC~3227 is somewhat uncertain due to spectral decomposition issues, and due to geometric rectification issues during the data reduction (J. Walsh, private communication). Except the central measurement, the observed $\oi/\sii$ agree with the RPC calculations. We emphasize that there are no free parameters in the RPC model calculations presented in Fig. \ref{fig: SII vs r}.

The \sii\ doublet is commonly used to measure $n$ (e.g. \citealt{Walsh+08}). In the lower panels of Fig. \ref{fig: SII vs r}, we compare the observed ratios of the \sii\ doublet with the RPC calculation. 
In NGC~3227, the observed $\sii$ ratio vs. $r$ relation is somewhat flatter than expected from RPC, and in NGC~4151, the observed $\sii$ ratios are typically higher than expected. \cite{Kraemer+00} found a similar discrepancy between the observed \sii\ ratios and the $n$ required by the other emission lines (see their figs. 4 and 7). 
The \sii\ ratio is sensitive to $n$ only at $10^2<n<10^4\cm^{-3}$, half the dynamical range which can be probed by $\oi/\sii$. 
Therefore, the observed \sii\ ratio will be more sensitive to projection effects of emission from gas on larger scales, which may explain the discrepancy.

\subsection{Forbidden line profiles}\label{sec: low lledd profiles}
\renewcommand{\lledd}{\dot m}
\newcommand{\dvblr}{{v}_{\rm BLR}}

\begin{figure*}
\includegraphics{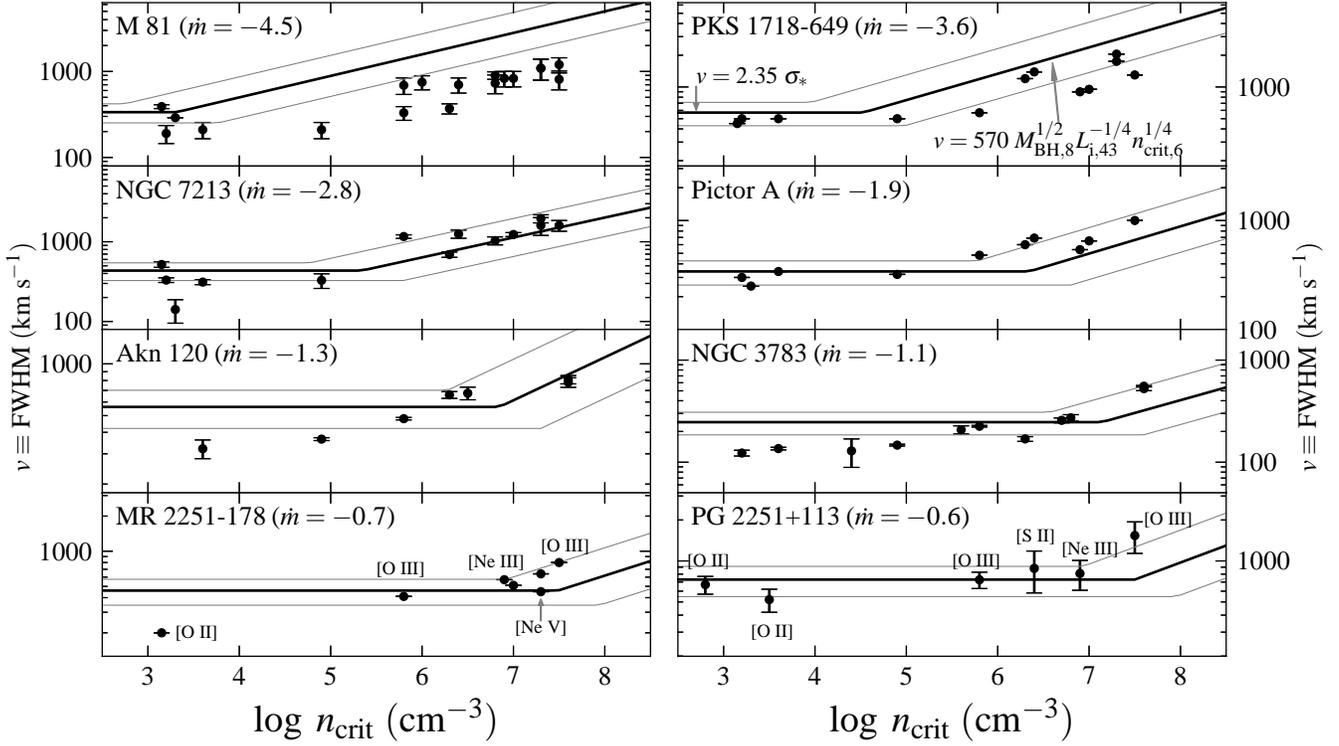}
\caption{The relation between line width and $\ncrit$ implied by RPC, compared to observations of type 1 AGN. 
The solid line in each panel denotes the expected $\dv(\ncrit)$, with no free parameters. The relation is flat at $\ncrit<\ncrit(\rgi)$ where the bulge dominates the gas kinematics, and has a slope of $1/4$ at $\ncrit>\ncrit(\rgi)$, where the kinematics are dominated by the black hole (eq. \ref{eq: FWHM vs ncrit}). The $\ncrit(\rgi)$ depends on $\lledd$ (eq. \ref{eq: rgi pressure}), which is noted in each panel. 
The normalization at $n>\ncrit(\rgi)$ is based on estimates of $\mbh$ and $\lion$ taken from the literature, with a factor of three error assumed on the $\mbh$ estimates (gray lines). 
The error bars denote the observed $\dv$ of different forbidden emission lines, from the references listed in Table \ref{tab: liners}. Some line designations are noted in the lower panels. 
With increasing $\lledd$, there is a clear increase in the lowest observed $\ncrit$ where the $\dv$ rises above $2.35\sV$, as expected from RPC. The observed $\dv(\ncrit)$ relation is generally consistent with the slope and normalization implied by RPC. Therefore, the gas at $r<\rgi$ is likely RPC.
}
\label{fig: LINERs}
\end{figure*}

In \S\ref{sec: emissivity vs. r} we showed that if $-2 < \eta < 0$, the emission of a forbidden line will be dominated by gas which resides at $r$ such that $n_{\rm f}(r) \sim \ncrit$. Since $n_{\rm f}\propto r^{-2}$, we expect the line profile of forbidden lines with high enough $\ncrit$ to be dominated by gas at $r<\rgi$, where $\rgi$ is the gravitational radius of influence of the black hole. Such emission lines are expected to show an increase of profile width with $\ncrit$. In contrast, a constant width is expected from lower $\ncrit$ lines, which originate at $r>\rgi$ where the gas kinematics are dominated by the bulge (\citealt{FilippenkoSargent88}; \citealt{Laor03}). In this section, we show that RPC and the $-2 < \eta < 0$ assumption is consistent with the observed widths of high $\ncrit$ forbidden lines in AGN.

At $r>\rgi$, we expect the emission line profile to be determined by the stellar velocity dispersion, $\sV$. Therefore, for a Gaussian profile the Full Width Half Max ($\dv$) is $2.35\sV$. 
At $r<\rgi$, assuming Keplerian motion and neglecting projection effects (see \citealt{Laor03}), we expect $\dv^2 = G\mbh/r$, where $\mbh$ is the black hole mass. Therefore, $\rgi$ can be derived from:
\begin{equation}
 \sqrt{\frac{G\mbh}{\rgi}} = 2.35 \sV
\end{equation}
Using $\sV/200 \kms = (\mbh/10^{8.12} \msun)^{1/4.24}$ from \cite{Gultekin+09},  we get
\begin{equation}\label{eq: rgi}
 \rgi = \frac{G\mbh}{5.5\sV^2} = 2.3~ M_{\rm BH, 8}^{\sim1/2} \pc
\end{equation}
where $\mbh = 10^8M_{\rm BH, 8}\msun$. Therefore, from eqs. \ref{eq: n vs r} and \ref{eq: rgi} 
\begin{equation}\label{eq: rgi pressure}
 n_{\rm f}(\rgi) = 1.6 \times 10^{8}~\lledd T_{f,4}^{-1}  \cm^{-3}
\end{equation}
where $\lledd$ is $\lbol$ in units of the Eddington luminosity, and we set $\lion=0.35~\lbol$, as above. 
At $r<\rgi$, we assume $\dv$ is dominated by emission from $r$ such that $n_f(r)=\ncrit$. Using eq. \ref{eq: n vs r} with $T_{f, 4}=1$ we get
\begin{equation}\label{eq: FWHM vs ncrit}
\dv =\sqrt{\frac{G\mbh}{r(\ncrit)}}=570~M_{\rm BH,8}^{1/2}~L^{-1/4}_{\rm i, 43}~n_{\rm crit,6}^{1/4}\kms
\end{equation}

\begin{table}
\begin{tabular}{l|c|c|c|c|c}
Object & $M_{\rm BH}$ & $L_{\rm i,43}$ & $\sV$ & Ref. & estimated $\mbh$ \\
\hline
M~81         & 7.8 & 0.01 & 143 & 5 & 6.8 \\
PKS~1718-649 & 8.5 & 0.3  & 243 & 2 & 8.0 \\
NGC~7213     & 8.0 & 0.6  & 185 & 1 & 8.2 \\
Pictor~A     & 7.5 & 2    & 145 & 2 & 7.9 \\ 
NGC~3783     & 6.9 & 3    & 105 & 3 & 7.4 \\ 
Ark~120      & 8.4 & 60   & 239 & 3 & 8.4 \\ 
MR~2251-178  & 8.1 & 100  & 196 & 2 & 8.4 \\ 
PG~2251+113  & 8.9 & 900  & 311 & 4 & 9.5 \\
\end{tabular}
\caption{
Properties of the objects in Fig. \ref{fig: LINERs}, compiled from the literature. The $\mbh$, $\lion$, and $\sV$ are given in units of $\log\ \msun$, $10^{43}\ergs$ and $\kms$, respectively. The references for the NLR $\dv$ measurements are: 
1. Fillipenko \& Halpern (1984)
2. Fillipenko (1985)
3. Appenzeller \& Oestreicher (1988)
4. Espey et al. (1994)
5. Ho et al. (1996).
The estimate of $\mbh$ based on RPC is described in \S\ref{sec: mbh measurement}. 
}
\label{tab: liners}
\end{table}

Fig. \ref{fig: n vs r} shows $\rgi$ for $M_{\rm BH, 8}=1$. 
The $\mbh$ will increase the width of an emission line if the intersection of the appropriate $\ncrit$ (dotted line) with the appropriate $\lion$ (solid line) are at $r<\rgi$. For example, \oiii\ $\lambda4363$ ($\ncrit=10^{7.5}\cm^{-3}$) will be emitted at $r<\rgi$ in $\lledd < 0.2$ AGN, while \oiii\ $\lambda5007$ ($\ncrit=10^{5.8}\cm^{-3}$) will have emission from $r<\rgi$ if $\lledd < 10^{-2.5}$. 

Figure \ref{fig: LINERs} compares the observed $\dv$ vs. $\ncrit$ with eq. \ref{eq: FWHM vs ncrit}, for the seven type 1 AGN with measurements of $v$ listed in \cite{Espey+94}. We add also M~81, which has measurements of $v$ of 18 forbidden lines in \cite{Ho+96}. We avoid type 2 AGN where high $n$ gas near the nucleus may be obscured. 
The estimates of $\mbh$, $\lion$, and $\sV$ for the different objects are gathered from the literature, as detailed in Appendix \ref{app: low lledd measurements}. They are listed in Table \ref{tab: liners}, together with the references for the $\dv$ measurements. We assume a factor of three uncertainty in the $\mbh$ estimates, and an uncertainty of $25\%$ in the $\sV$ estimates. The uncertainty in $\lion$ in the Seyferts is small compared to the uncertainty in the $\mbh$ estimate, however the estimate of $\lion$ in the LINERs (M~81, PKS~1716-649 and NGC~7213) is highly uncertain. 

Note that some of the emission lines in Fig. \ref{fig: LINERs} have high IP, implying that they are emitted from a layer in the slab in which $n<n_{\rm f}$. The highest IP lines shown, \nev\ and \fevii\ (IP$=97\eV$), have an emissivity averaged $n\approx0.1~n_{\rm f}$ in our \cloudy\ models. 
Therefore, the $r$ where $n=\ncrit$ in these lines is smaller by a factor of $0.1^{1/2}$ than the $r$ derived assuming $n_f=\ncrit$. Hence, the observed $\dv$ is expected to be larger by a factor of $\sim 0.1^{-1/4}$ than expected from eq. \ref{eq: FWHM vs ncrit}. For simplicity, and due to the uncertainties induced by the unknown $\Omega(r)$ distribution, the assumption of Keplerian motion, and possible projection effects, we do not incorporate this additional complication in our calculations. 

The objects in Fig. \ref{fig: LINERs} span a dynamical range of $10^4$ in $\lledd$. With increasing $\lledd$, there is a clear increase in the observed $\ncrit$ where the $\dv$ rises above $2.35\sV$, as expected from eq. \ref{eq: rgi pressure}.
The slope of the observed relation between $\dv$ and $\ncrit$ at $\ncrit>\ncrit(\rgi)$ is consistent with $1/4$, as expected from eq. \ref{eq: FWHM vs ncrit}. With the exception of M~81, the actual observed $\dv$ are generally consistent with eq. \ref{eq: FWHM vs ncrit} and the $\mbh$ and $\lion$ estimates. Therefore, the gas at $r<\rgi$ is likely RPC. We emphasize again that there are no free parameters in the RPC results. 

It is also possible to estimate $\dv(\ncrit)$ directly from the FWHM of the broad \Ha, $\dvblr$, with no relation to the value of $\mbh$. In a single object, the low ionization part of the BLR appears to be dominated by gas from a small range of $r$, which satisfies $r_{\rm BLR} \propto L_{\rm 1450\AA}^{0.55}$ (\citealt{Kaspi+05}). Therefore, assuming a Keplerian velocity field and using the Kaspi et al. relation, we get
\begin{equation}\label{eq: Dv BLR}
 \frac{\dv(\ncrit)}{\dvblr} = \left(\frac{r(\ncrit)}{r_{\rm BLR}}\right)^{-1/2} =
  \left(\frac{1.4~n_{\rm crit,6}^{-1/2}~L_{\rm i,43}^{1/2}\pc}{0.0036 ~L_{\rm i,43}^{\sim1/2}\pc}\right)^{-1/2} 
= 0.05 ~ n_{\rm crit,6}^{1/4} 
\end{equation}
where we assumed $\lbol/L_{\rm 1450\AA}=4$ (\citealt{Richards+06}).

\subsection{Emission line ratios: high IP vs. low IP}\label{sec: high IP vs low IP}

The RPC slab structure seen in Figs. \ref{fig: profile vs tau}--\ref{fig: different ngamma} implies a highly ionized surface followed by a less ionized inner layer. Hence, in RPC gas the high IP and low IP emission lines come from the same slab, and their expected emissivity ratios can be calculated. This predictability is distinct from other models, such as locally-optimal emitting clouds (\citealt{Ferguson+97}), where lines with different IP come from different slabs, and therefore their emission ratios are not constrained. 

G04 and \cite{Gorjian+07} showed that the observed unresolved $\nev\ 3426\AA/\neiii\ 3869\AA$ and $\nev\ 14.32\mic\ /\neiii\ 15.55\mic$ are generally consistent with dusty RPC models with $n_{\rm f}=1000\cm^{-3}$. 
We extend their analysis, by comparing observed emission line ratios from well defined samples, with RPC models with $n_{\rm f}$ in the range $30-3\times10^7\cm^{-3}$. Also, we compare the RPC calculations with resolved observations.

\subsubsection{Unresolved observations}

\begin{figure*}
\includegraphics{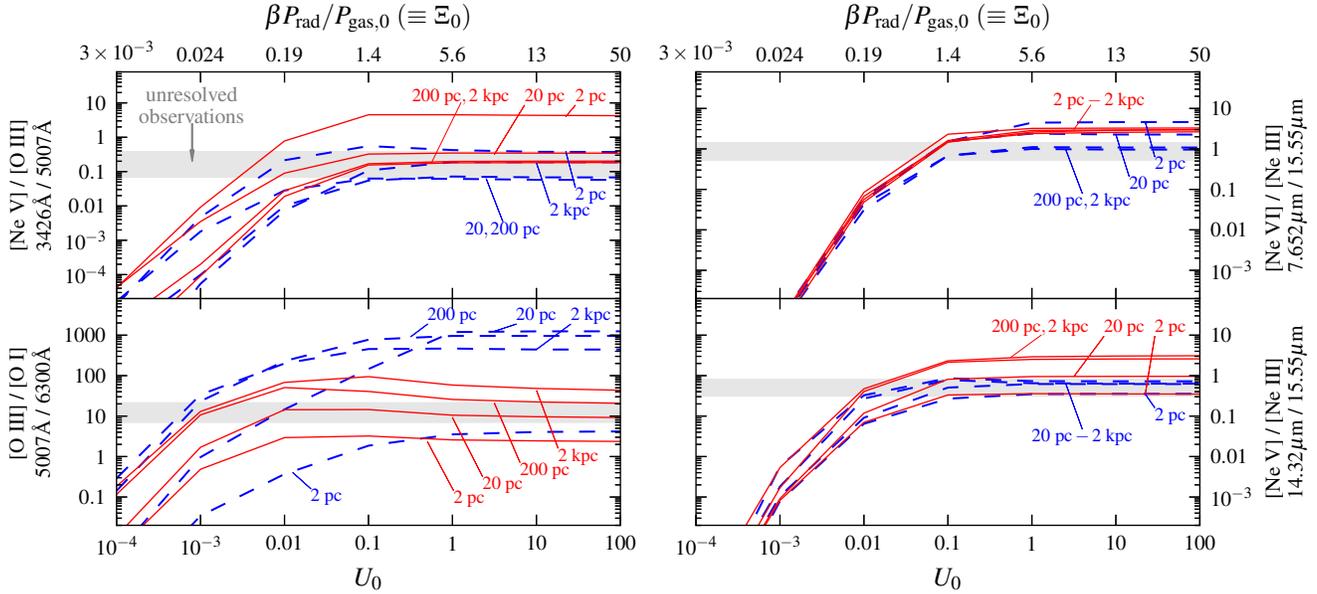}
\caption{Expected vs. observed high ionization to low ionization emission line ratios. 
In each panel, the four solid red lines represent dusty models with $Z=2\zsun, \aeuv=-1.6$ and $r=2-2000~L_{\rm i,45}^{1/2}\pc$. 
The four dashed blue lines represent dust-less models with the same parameters. 
For each $r$, models are calculated for a range of $U_0$, or equivalently, a range in $\Xi_0$ (noted on top). 
Models with $U_0>>0.03$ ($\Xi_0>>1$) are RPC, and their emission line ratios are independent of $U_0$. Models with lower $U_0$ require an additional confinement source which is stronger than RPC. 
Gray stripes show the one-sigma range of emission line ratios in unresolved observations of luminous type 1 AGN.
The observed ratios are consistent with the RPC calculation to within a factor of $2-3$, and rule out a significant contribution to these emission lines from slabs with $U_0<<0.03$. 
The observed $\oiii/\oi$ suggests that the layer which emits these lines is dusty.
}
\label{fig: U dependence}
\end{figure*}

We choose emission line couples according to the following guidelines:
\begin{enumerate}
 \item The lines differ in IP, so the luminosity ratio is sensitive to the relative emission from different layers in the slab. 
 \item Strong emission, so the emission lines are observable with high S/N.
 \item The lines have similar $\ncrit$, to reduce the dependence of the luminosity ratio on the unknown $\Omega(r)$.
 \item The lines are weak in star forming regions, to avoid contamination from other sources.
 \item The lines are not blended with broad lines or stellar absorption features, so the reliability of the measurements is high. 
 \item Lines from a noble element are preferred, so the luminosity is not directly sensitive to depletion.
\end{enumerate}

The chosen couples are the optical $\nev\ 3426\AA/\oiii\ 5007\AA$ and $\oiii\ 5007\AA / \oi\ 6300\AA$, and the IR $\nevi\ 7.652\mic / \neiii\ 15.55\mic$ and $\nev\ 14.32\mic\ /\neiii\ 15.55\mic$. 
The differences in $\ncrit$ of these four couples are 1.5, 0.5, 0. dex, and 0.8 dex, respectively. 
The IPs are 0, 35, 41, 97 and 126\eV, for \oi, \oiii, \neiii, \nev, and \nevi, respectively. 
We use emission line observations of luminous type 1 AGN with well-defined selection criteria, as detailed in Appendix \ref{app: emission line ratios}. Type 1 AGN are preferred since the narrow line ratios are not part of the selection process, and reddening effects should be less severe than in type 2 AGN. 

The predicted emission line luminosity ratios are calculated with \cloudy\ (\S\ref{sec: cloudy}), using dusty and dust-less models with $\aeuv=-1.6,\ Z=2\zsun$ and $r$ in the range $2-2000~L_{\rm i,45}^{1/2}\pc$. The observed emission in each object is expected to be a weighted sum of the emission from slabs at different $r$. 
To emphasize the effect of RPC, we vary $U_0$ from $10^{-4}$ to $100$ in one dex intervals.
Figure \ref{fig: U dependence} compares the calculation of the models with the observed values. 
For each $U_0$ we note on top the appropriate $\Xi_0$ for the dusty model with $r=20\pc$. Other models have $\Xi_0$ which are offset from the noted value by $<0.4$ dex. 
Models with $U_0>>0.03$ ($\Xi_0>>1$) are RPC, and their emission line ratios are independent of $U_0$, as expected from Figs. \ref{fig: different P0} and \ref{fig: different P0 no dust}. 
Models with $U_0<<0.03$ have $\prad<<\pgas$, and therefore require an additional confinement mechanism which is stronger than RPC. 
The uncertainty in the expected ratios due to a possible factor of two in $Z$ or a change of $\pm0.2$ in $\aeuv$ is $0.5$ dex for $\oiii/\oi$ and $0.2$ dex for the three other ratios. 

In the $\nev/\oiii$ panel, the calculations of all RPC models except the dusty $r=2\pc$ model are within the observed range of values. 
Therefore, given the uncertainties mentioned above, the observed $\nev/\oiii$ are consistent with a dust-less RPC model with any distribution in $r$, and also with a dusty RPC model, as long as the emission of these two lines is not dominated by gas at $2~L_{\rm i,45}^{1/2}\pc$. 

Note that $\oiii$ is efficiently emitted from gas with $U_0$ as low as $\approx10^{-3.5}$, so in principle gas with $U_0<<0.03$ can contribute significantly to the observed \oiii. However, the fact that $U_0<<0.03$ models underpredict the observed $\nev/\oiii$ by orders of magnitude, suggest that it is unlikely that such gas dominate the $\oiii$ emission.

In the $\nevi/\neiii$ panel, the dust-less RPC models with $r=200\pc$ and $r=2\kpc$ are within the observed range of ratios, while the $r=20\pc$ and $r=2\pc$ models are above the mean observed value by a factor of three and five, respectively. Note however that for a broad distribution in $r$, a slab at $r=2\pc$ will not affect the observed $\nevi/\neiii$ significantly because both lines are collisionally de-excited (Fig. \ref{fig: ew vs r}).  All dusty RPC models are above the mean observed value by a factor of $3.5$. Therefore, the observed $\nevi/\neiii$ suggest either a dust-less RPC model, or a dusty RPC model with additional contribution to $\neiii$ from $U_0<<0.03$ slabs, which decreases the observed $\nev/\neiii$ ratios from the pure-RPC value. However, given the uncertainty mentioned above, the pure-RPC dusty models cannot be ruled out. A similar behavior is observed in the $\nev/\neiii$ panel, where dust-less RPC models are consistent with the observed values, while dusty RPC models with $r=20\pc-2\kpc$ overpredict the observed mean value by a factor of $1.5 - 5$. 

In the $\oiii/\oi$ panel, the dusty RPC models span the entire range of observed values, and therefore the observed $\oiii/\oi$ are consistent with a dusty RPC model with slabs from a broad distribution in $r$. The RPC dust-less models with $r\geq 20\pc$ overpredict $\oiii/\oi$ by a factor of $30-100$. This large difference between the dusty and dust-less models is because the $\oiii$ emissivity decreases due to absorption of ionizing photons by the dust, while the \oi\ emission increases because of the photoelectric heating of the gas by the grains. In the $r=2\pc$ dust-less model, the calculated \oi\ emission per unit $\Omega$ is enhanced by a factor of 50 compared to models with larger $r$. It is not clear whether this huge increase in \oi\ emission is a physical effect, or some artifact of the calculation. Therefore, this panel suggests that the gas is dusty, somewhat in contrast with the conclusion from the other panels. In \S\ref{sec: dust destruction}, we present additional evidence that the \oiii\ 
and \oi\ emitting layers are likely dusty, while in the layers which emit \nev\ and \nevi\ the dust is at least partially destroyed. 

To summarize, the RPC calculations are consistent to within a factor of a few with the observations of unresolved emission line ratios, despite the small dynamical range of emission line ratios permitted by the RPC models.

\subsubsection{Resolved observations -- $\neiii/\nev$}\label{sec: Mazzalay}

\cite{Mazzalay+10} compared $\neiii\ \lambda3426/\nev\ \lambda3869$ with $r$ in nine local Seyferts (their fig. 20). 
Fig. \ref{fig: ew vs r} implies that in all the off-center observations of \citeauthor{Mazzalay+10}, which are at $r>10\pc$, the value of $\neiii/\nev$ is not expected to change with $r$ by more than a factor of two\footnote{None of the \cite{Mazzalay+10} objects is likely to have $\lion>10^{46}\ergs$.}. 
Indeed, Mrk~573, NGC~4507, Mrk~348, NGC~7682, NGC~5643 and NGC~3081 show that $\nev/\neiii$ vs. $r$ is constant up to a factor of about two. In contrast, in non-RPC models, $\neiii/\nev$ drops by a factor of 1000 between $U_0=0.1$ and $U_0=10^{-3}$, similar to the drop in the $\nev\ 14.32\mic/\neiii\ 15.55\mic$ ratio seen in the top right panel of Fig. \ref{fig: U dependence}. Therefore, non-RPC models will have difficulty explaining why $U_0$ is so constant at different $r$. 

The observed values of $\neiii/\nev$ in these six objects is in the range $0.4 - 3$, compared to $\neiii/\nev=0.4-0.8$ expected in dusty RPC models with $1<Z/\zsun<4$ and $-1.6<\aeuv<-1.2$, and $\neiii/\nev=0.5-3$ expected in dust-less RPC models with the same range in $Z$ and $\aeuv$. Therefore, both the lack of trend of $\neiii/\nev$ with $r$, and the observed values of $\neiii/\nev$, suggest that the gas which emits \neiii\ and \nev\ in these six objects is RPC. In the other objects, however,  some of the observed ratios, at some specific positions, can deviate significantly from RPC, which may indicate non-RPC conditions. 

In the six objects with constant $\nev/\neiii$ vs. $r$, 
one finds $\oii/\oiii= 0.06-0.4$. 
For comparison, dusty RPC models with the range of parameters noted above give $\oii/\oiii=0.08-0.25$ in the low $n_{\rm f}$ limit. Reddening along the line of sight can decrease $\oii/\oiii$, while star formation will increase $\oii/\oiii$. 
While the observed $\oii/\oiii$ are comparable to the expected values at the low $n_{\rm f}$ limit, the dispersion per object is larger than expected from a pure-RPC model, and the expected decrease in $\oii/\oiii$ with decreasing $r$ is not seen. 
The fact that $\oii/\oiii$ does not decrease apparently contradicts RPC. 
Possibly, projection effects (see \S\ref{sec: resolved}) increase the apparent $\oii/\oiii$ to the low-$n$ value. 
This conjecture can be tested using the \oii\ line width. If the projected $r$ is $<\rgi$, but the \oii\ emission comes from larger $r$, then \oii\ should not show the expected increase in line width.

\subsubsection{Resolved observations -- $\loiii/\lx$}\label{sec: Bianchi}

\cite{Bianchi+06} showed that the \Chandra\ maps of extended $\lx$ overlap the \HST\ maps of $\loiii$ in eight Seyfert 2s, selected from the FIR-bright \cite{Schmitt+03} catalog based on the availability of a \Chandra\ observation. Spectroscopy showed that $\lx$ is dominated by emission lines, which likely arise in photoionized gas. Bianchi et al. used this overlap to show that $U$ is independent of $r$, and therefore $n \propto r^{-2}$. As can be seen in Fig. \ref{fig: ew vs r}, this nearly constant $\loiii/\lx$ is a direct consequence of RPC, under the condition that $n_{\rm f}<<\ncrit(\oiii)$, which is satisfied for the Bianchi et al. 
observations\footnote{
Fig. \ref{fig: ew vs r} shows that $n_{\rm f}=\ncrit(\oiii)=10^{5.8}\cm^{-3}$ at $r=16~L_{{\rm i,} 45}^{1/2}\pc$, while the extended emission in the \cite{Bianchi+06} maps is on scales $>>25\pc$. None of the \cite{Bianchi+06} objects is likely to have $\lbol>>10^{45.5}\ergs$.}.
In other words, RPC gives a physical interpretation for the $n\propto r^{-2}$ relation found by Bianchi et al. 

The observed $\loiii/\lx$ in the \cite{Bianchi+06} objects are 2.8 -- 4.8, except NGC~7212 which has $\loiii/\lx=11$. In the low-$n$ limit, the dusty $Z=2\zsun$ RPC models give $\loiii/\lx=$1.4, 5.7, and 23 for $\aeuv=$ --1.2, --1.4, and --1.6, respectively. Increasing or decreasing $Z$ by a factor of two changes $\loiii/\lx$ by $<40\%$. 
Therefore, the observed $\loiii/\lx$ are in the range of $\loiii/\lx$ derived from the RPC models using reasonable values of $\aeuv$.

\begin{figure*}
\includegraphics{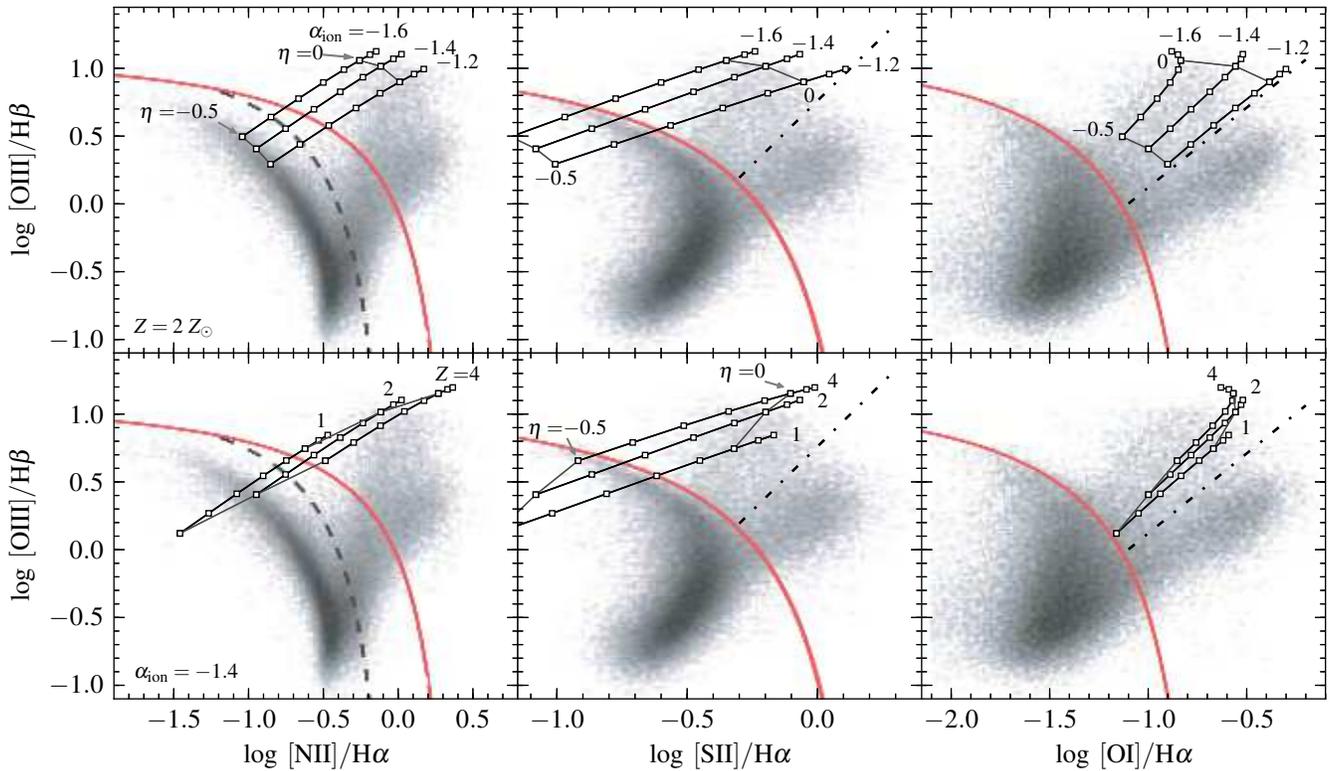}
\caption{
Expected BPT ratios for an ensemble of RPC slabs with a distribution in $r$, compared to observations of SDSS galaxies. 
Each black line marks the RPC calculations, for $Z=2\zsun$ and different $\aeuv$ (top panels, $\aeuv$ noted), or for $\aeuv=-1.4$ and different $Z$ (bottom panels, $Z$ noted). The white squares mark different $\eta$, where $\eta$ is the index of the slab covering factor distribution as a function of $r$ (eq. \ref{eq: eta}). The $\eta$ vary between $-0.5$ and $0.2$ in steps of 0.1. Same-$\eta$ models with $\eta=0$ or $\eta=-0.5$ are connected by a thin gray line, and $\eta$ is noted. Lower $\eta$ produce lower ratios of forbidden lines to recombination lines, due to the collisional de-excitation of the forbidden lines at small $r$. 
The background shows the observed emission line ratios in SDSS emission line galaxies (fig. 1 in K06), and the various classification lines from Kewley et al. (2001), Kauffmann et al. (2003), and K06. Seyferts reside above the solid red and black dash-dotted lines. 
The RPC model with $\aeuv\lesssim-1.4$, $Z\gtrsim2\zsun$ and $\eta\sim-0.1$ reproduces the observed BPT ratios at the high-\oiii/\Hb\ end of the Seyfert distribution. 
All of these parameters are expected from independent observations. 
An additional star forming component is required to explain the entire distribution of BPT ratios in SDSS Seyferts.
}
\label{fig: BPT}
\end{figure*}

\subsection{Emission line ratios: BPT}
\newcommand{\ngmax}{n_{\gamma, {\rm max}}}
\newcommand{\rbreak}{r_{\rm break}}

G04 calculated the BPT ratios (\citealt{Baldwin+81}; \citealt{VeilleuxOsterbrock87}) of dusty models with $n_f=10^2-10^4\cm^{-3}$. The G04 models include both RPC models with $U_0>>0.03$, and models with $U_0<<0.03$ which are effectively constant-$n$ models. 
\citeauthor{Kewley+06} (2006, hereafter K06) compared the G04 models with SDSS Seyferts in the $\oiii/\Hb$ vs. $\oi/\Ha$ BPT diagram (fig. 23 there). The shown range of emission line ratios implied by the different $U_0$ is basically of non-RPC models, since the RPC models all converge to a single solution, as noted in G04. K06 found that the BPT ratios of RPC slabs are consistent with the high-$\oiii/\Hb$ end of the observed distribution in SDSS Seyferts.

Here, we extend the K06 analysis to a distribution of slabs with $10^{-1.5}\leq\ng\leq10^{7.5}\cm^{-3}$, implying $1<n_{\rm f}<10^9\cm^{-3}$. We use only RPC models ($U_0>>0.03$), where the emission line ratios are independent of $U_0$. Slabs with $\ng<1\cm^{-3}$ are disregarded since the ambient ISM pressure will likely dominate the radiation pressure (Fig. \ref{fig: n vs r}), and such slabs will not be RPC. This minimum $\ng$ corresponds to $r_{\rm out}=1.2~L_{\rm i,43}^{1/2}\kpc$ (eq. \ref{eq: ng}). Slabs with $\ng\geq 10^{8.5}\cm^{-3}$ are disregarded because they are part of the BLR\footnote{As mentioned above, our models do not include the expected sublimation of small grains at $\ng>10^{5.9}\cm^{-3}$.}. 
We assume a power law distribution with index $\eta$ (eq. \ref{eq: eta}), and sum the emission from different slabs weighted by the implied $d\Omega$. 

Note that in type 2 AGN the high $\ng$ gas near the center may be obscured, and therefore should not enter the calculation, though at which $\ng$ this occurs is not well-constrained. In the BPT ratios, decreasing $\eta$ and lowering the maximum $\ng$ is degenerate, so the $\eta$ derived below may be somewhat overestimated. 
However, the similarity of the BPT ratios in type 1 and type 2 AGN which are selected similarly (\citealt{SternLaor13}) implies that obscuration does not play a significant role in the BPT ratios. 

Figure \ref{fig: BPT} compares the RPC calculations with the emission line measurements of SDSS galaxies (fig. 1 in K06). We add several commonly used classification lines: the theoretical classification lines from \cite{Kewley+01} which separate star forming (SF) galaxies and AGN (red solid lines), the empirical classification line from \cite{Kauffmann+03}, which separate pure-SF and SF-AGN `Composites' (dashed line in the $\nii/\Ha$ panel), and the K06 empirical separation between Seyferts and LINERs (dash-dotted lines in the $\sii/\Ha$ and $\oi/\Ha$ panels). 
Each black solid line represents the RPC calculation for $-0.5\leq\eta\leq0.2$, with white squares representing steps of $0.1$ in $\eta$. The top panels show models with $Z=2\zsun$ and several reasonable values of $\aeuv$, while the bottom panels show models with $\aeuv=-1.4$ and several reasonable values of $Z$. 
Lower $\eta$ produce lower BPT ratios, due to collisional de-excitation of the forbidden lines in slabs with small $r$, which are more significant at low $\eta$. 

RPC models with $Z=2\zsun$, $\aeuv=-1.4$, and $\eta\approx-0.1$ reproduce the observed BPT ratios at the high-\oiii/\Hb\ end of the Seyfert distribution. Models with $\eta\approx-0.1$ and $Z=4\zsun$ or $\aeuv=-1.2$ are also generally within the observed range of values, though they overpredict the observed \oi/\Ha\ in most Seyferts by $\sim 0.2$ dex. The $\eta\approx-0.1$ implied by the BPT diagrams is consistent with the $\eta \sim 0$ expected from the flat IR slope in the mean quasar SED (\S\ref{sec: emissivity vs. r}). Therefore, the conclusion of K06 mentioned above applies also when summing slabs with a broad distribution of $n$. As noted by K06, an additional SF component is required to reproduce the entire observed distribution of BPT ratios in SDSS Seyferts.

\subsection{The extent of RPC in type 2 quasars}

Recently, \cite{Liu+13a} resolved $\oiii/\Hb$ in $z\sim0.5$ type 2 quasars
selected from the \cite{Zakamska+03} sample, with a resolution of $\sim
3\kpc$.
They find a constant $\oiii/\Hb = 12.3\pm2.7$ extending out to $r=\rbreak$,
with $4 < \rbreak < 11\kpc$. At $r>\rbreak$, $\oiii/\Hb$ declines with $r$.
In RPC, a constant $\oiii/\Hb=13$ is expected at $r>300~L^{1/2}_{\rm i,45}
\pc$ (Fig. \ref{fig: ew vs r}), consistent with the observed $\oiii/\Hb$ at
$r<\rbreak$ and the $\lion = 10^{46}-10^{47}\ergs$ of the
\citeauthor{Liu+13a} sample\footnote{We derived $\lion$ from the $\loiii$
listed in \cite{Liu+13b} and the $\lbol$ vs. $\loiii$ relation of
\cite{SternLaor12b}.}.

The decrease in $\oiii/\Hb$ at $r>\rbreak$ may indicate that the ambient pressure
$\pgasz$ exceeds $\prad$, causing $U_0$ to decrease with increasing $r$.
Therefore, $\rbreak$ is the maximum $r$ where RPC is applicable. Since the
value of $r$ where $\pgasz=\prad$ depends on $\lbol$ and on $\pgasz$, we
expect a correlation between $\rbreak$ and $\lbol$.
Indeed, $\rbreak$ and $\loiii$ have a Pearson correlation coefficient of 0.45
in the \citeauthor{Liu+13a} sample, with a 12\% chance of random occurrence.
Hence, there is a possible relation between $\loiii$ and $\rbreak$, as
expected from RPC, though more data is required for a definitive answer. The
implied $\pgasz(\rbreak)$ is $220\pm98\cm^{-3}$ (in units of
$2.3\cdot10^4\kB$).

\section{Dust existence}\label{sec: dust destruction} 

In Fig. \ref{fig: different P0 no dust} we showed that dust has a significant effect on the RPC slab structure. 
In this section, we present the relevant theoretical considerations and observational evidence for the existence of dust in ionized gas in AGN.

\subsection{Theoretical considerations}
\newcommand{\vd}{v_{\rm drift}}
\newcommand{\cs}{c_{\rm s}}

\subsubsection{Grain sublimation}\label{sec: dust sublimation}

Fig. 8 in \cite{LaorDraine93} shows the values of $r \lbol^{-1/2}$ where grains with different compositions and sizes sublimate.
Assuming $\hnu=36\eV$ and $\lion=0.35\lbol$, as above, Silicate grains with radii $a=0.005, 0.25$ and $10\mic$ sublimate at $\ng = 10^{5.9}, 10^{7.3}$ and $10^{8.3}\cm^{-3}$, respectively. Graphite grains with the same $a$ sublimate at $\ng = 10^7, 10^{8.6}$ and $10^{8.6}\cm^{-3}$. 
Therefore, at $\ng > 10^{8.6}\cm^{-3}$ all dust grains sublimate. At $10^{5.9} < \ng < 10^{8.6} \cm^{-3}$, the dust content strongly depends on $\ng$.

\subsubsection{Grain sputtering}
\newcommand{\qpr}{\langle Q_{\rm pr} \rangle}
\newcommand{\qprav}{\langle \overline {Q_{\rm pr}} \rangle}

Another source of grain destruction is sputtering due to collisions with gas particles. The sputtering efficiency depends on the relative velocity between the grain and individual gas particles. This velocity is either the sound speed $\cs$, or the dust drift velocity in the gas rest frame $\vd$, if the drift is supersonic. 
For stationary grains sputtering is efficient at $T>10^5-10^6\K$, or $c_{\rm s}>40-130\kms$ (\citealt{Draine11b}). Most line emission in the RPC slab occurs at lower $T<<10^5\K$ (see appendix \ref{app: ionization structure}), so without $\vd>>\cs$, sputtering is unlikely to have a significant effect on the dust content in the line-emitting layer of the slab. Accurate calculation of $\vd$ in AGN has not been performed yet, and is beyond the scope of this work. 
However, it is relatively straightforward to derive an upper limit on $\vd$, so it is possible to understand in which layers sputtering is possible, and in which layers it is unlikely. 

The terminal $\vd$ of a neutral grain can be derived by balancing the radiation force on the grain with the force of collisional drag (e.g. \citealt{DraineSalpeter79}). In the limit of highly supersonic drift, 
\begin{equation}\label{eq: drag}
 \frac{\lbol}{4\pi r^2 c} \qpr\pi a^2 = \pi a^2 n \mp \vd^2
\end{equation}
where $\qpr\pi a^2$ is the radiation pressure cross section averaged over the incident spectrum (\citealt{Draine11b}). 
Replacing $\lbol/(4\pi r^2 c)$ with $1.5\beta\prad$, appropriate for the $\beta=1.9$ derived above and the $\lion=0.35~\lbol$ in our assumed SED, we get
\begin{eqnarray}\label{eq: vd}
1.5\qpr\beta \prad = n \mp \vd^2  = n\mp\vd^2 \left(\frac{2 K T }{\mp c_{\rm s}^2}\right) = \pgas \left(\frac{\vd}{c_{\rm s}}\right)^2 \nonumber \\
\Rightarrow  \left(\frac{\vd}{c_{\rm s}}\right)^2 = 1.5\qpr\frac{\beta \prad}{\pgas} = 1.5\qpr\Xi = 1.5\qpr\tau^{-1}
\end{eqnarray}
where we used eq. \ref{eq: xi vs tau} in the last equality, which is valid at $\tau<1$. 
Charged grains will also experience Coulomb drag, which may decrease $\vd$.

The value of $\qpr$ is unlikely to be larger than $\sim2$ (e.g. \citealt{LaorDraine93}). 
Therefore, at $\tau=0.3$, where $T=30,000\K$ (Fig. \ref{fig: profile vs tau}), 
we find $\vd\leq(1.5\cdot2\cdot\tau^{-1})^{1/2}\cs=70\kms$, and sputtering may be efficient in destroying the grains. 
The dependence of $\vd/\cs$ on $\qpr$, together with the dependence of the sputtering efficiency at a certain $T$ on grain properties, can lead to a situation where only part of the dust is destroyed in this layer.
Near the ionization front, where $\Xi\approx1$ and $T\approx 10^4\K$, sputtering is highly ineffective.

\subsection{Observational constraints}

As noted above, dust survival depends on the depth within the slab. We therefore divide the slab into two different layers, and analyze the observational evidence for dust existence in each of them.

\subsubsection{Inner layer {\rm (IP $\lesssim40\eV$)}}

The layer which emits \oiii\ and other lines with similar or lower IP occurs at $\tau\gtrsim1.5$ (app. \ref{app: ionization structure}), where dust grains will likely survive. 
There are several indications that this inner layer of the NLR is indeed dusty. First, \cite{Galliano+03, Galliano+05} found that the extended MIR emission in NGC 1068 is well correlated with the \oiii\ emission, suggesting that presence of dust grains in the line-emitting gas. 
Second, the observed $\oiii/\oi$ (lower left panel in Fig. \ref{fig: U dependence}) suggest a dusty RPC model. 
A third piece of evidence is the lack of detection of \caii\ 7291\AA\ in AGN, which suggests that Ca is highly depleted onto dust grains (\citealt{Ferland+93}; \citealt{Villar-MartinBinette97}; \citealt{Ferguson+97}; \citealt{Shields+99}; \citealt{Cooke+00}). The dust-less RPC model with $Z=2\zsun,\ \aeuv=-1.6$ and $\eta=0$ gives $\caii/\sii\ \lambda6716=0.11$. For comparison, the mean SDSS spectrum from \cite{VandenBerk+01} shows a prominent \sii\ feature, while \caii\ is not detected, suggesting $\caii/\sii<0.02$. Therefore, Ca is depleted by at least a factor of $\sim5$, implying the existence of dust in this low IP layer.

\subsubsection{Outer layer {\rm (IP $\gtrsim100\eV$)}}
Ne$^{4+}$ (IP $=97\eV$) appears at $\tau\sim0.3-1.5$ (app. \ref{app: ionization structure}). At the low $\tau$ end of this layer, sputtering may be efficient in destroying the grains (see above).
The \fevii\ $\lambda6087/\nev\ \lambda3426$ ratio has been suggested as a good tracer of the relative abundance of these two elements, due to the similar IP of the two ions (\citealt{NussbaumerOsterbrock70}). Therefore, the ratio of these two lines is a good measure of Fe depletion, which is depleted by a factor of 100 in the dusty ISM. 
\cite{VandenBerk+01} and \cite{Shields+10} found a mean $\fevii / \nev =0.3$ in SDSS quasars, while \cite{Nagao+03} found \fevii\ / \nev $=0.5\pm0.3$ in nearby type 1 AGN and $0.3\pm0.2$ in nearby type 2 AGN. 
For comparison, the RPC dusty model with $Z=2\zsun, \aeuv=-1.6$ and $\eta=0$ gives $\fevii\ / \nev = 0.012$.
while the dust-less model with the same parameters gives $\fevii\ / \nev= 0.55$. 
Therefore, the abundance of Fe, relative to the abundance of Ne, is much higher than the depleted abundance seen in the ISM. A similar conclusion was reached by \cite{Ferguson+97}, D02, \cite{Nagao+03} and \cite{Shields+10}.

The high abundance of Fe relative to the depleted abundance implies that if grains existed in this layer at some period, a non-negligible fraction of them has been destroyed. However, we note that even if 50\% of the dust has been destroyed, the Fe abundance would increase by a factor of 50, and thus be similar to the abundance in a dust-less model, while $\sigbar$ will decrease only by a factor of two, and thus the slab structure would be similar to the dusty models. For the dust opacity to decrease to $\sth$, 99.9\% of the dust needs to be destroyed (see eqs. \ref{eq: dusty sigbar}--\ref{eq: dust-less sigbar}). A selective destruction of grains is likely under some conditions due to the dependence of the destruction mechanisms on grain properties (see previous section).  Therefore, abundance measures are not robust ways to determine whether the dust opacity has been significantly reduced, and the question of whether some dust exists in this outer layer remains open.

\newcommand{\acl}{f_{\rm GMC}}
\newcommand{\astar}{f_{\rm star}}
\newcommand{\sigstar}{\sigbar_{\rm star}}
\newcommand{\reff}{r_{\rm eff}}
\newcommand{\tcomp}{T_{\rm C}}
\newcommand{\arad}{f_{\rm rad}}
\newcommand{\aother}{f_{\rm other}}

\section{Discussion and implications}\label{sec: implications}

\subsection{Validity of the RPC assumptions}

The RPC structure is calculated using the radiative force exerted by the ionizing radiation.
When does the radiative force dominate?
We define the radiative force per H-nucleus, normalized by $\mp$, as $\arad$.
At the illuminated surface of a dusty slab,
\begin{equation}\label{eq: frad dusty}
 \arad = \frac{\beta\prad \sigbar}{\mp} = 1.4 \times 10^{-4} L_{{\rm i,}45}r_{50}^{-2}\sigbar_{-21} \cm \s^{-2}
\end{equation}
The gas may reside in a typical giant molecular cloud (GMC). In this case the self-gravity force $\acl$ is
\begin{equation}\label{eq: fcl}
 \acl = 1.5 \times 10^{-8} \frac{\mcl}{10^5 \msun}\left(\frac{\dcl}{10\pc}\right)^{-2} \cm \s^{-2}
\end{equation}
where $\mcl$ and $\dcl$ are the the mass and size of the cloud, respectively.

Another force which may be significant in the ISM is radiation pressure from stellar light $\astar$.
Adopting $\prad=u/3$, for an isotropic radiation field, where $u=8.6\times 10^{-13}\ergs\cm^{-3}$ (\citealt{Draine11b})
is the energy density of the stellar radiation in the Galaxy,
gives
\begin{equation}\label{eq: fstar}
\astar=1.8\times 10^{-11} \cm \s^{-2}
\end{equation}
where we used $\sigbar=10^{-22}\cm^2$ which is the dust absorption cross section at the peak of the stellar
emission, around 1~$\mu m$. Clearly, the AGN radiative force dominates the
force from the ambient stellar light,
on all scales, as expected if the AGN is more luminous than the host galaxy.

For a GMC, we get that $\arad > \acl$ at $r < 4.8~ L_{{\rm i,}45}^{1/2}\kpc$.
Thus, once the AGN luminosity clearly dominates the host luminosity, i.e. $\lbol>10^{44}\ergs$,
the AGN radiative force can dominate the self-gravity of a GMC quite far out
on the host galaxy scale.
This force can compress the GMC, and possibly affect the star formation rate on large scales
in the host galaxy.

We note in passing that photoionized gas may be confined even when it is
optically thin and radiatively accelerated.
In the accelerated frame there will still be a differential acceleration, a
factor of $\tau$ smaller than for a static
slab, which will lead to a correspondingly smaller pressure gradient, and
therefore densities also a factor of $\tau$ smaller.
The structure of such a layer has been explored in various studies on AGN
(\citealt{Weymann76}; \citealt{ScovilleNorman95}; \citealt{CheloucheNetzer01}), and may be subject to various instabilities (\citealt{MathewsBlumenthal77}; \citealt{Mathews82, Mathews86}).

\subsection{Comparison with LOC}

The locally optimal emitting clouds model (LOC, \citealt{Ferguson+97})
suggests that the narrow line emission in AGN originates from an ensemble of
clouds with a distribution of $U$ and $n$, where each line originates from
the LOC which maximizes its emission. 
For comparison, the RPC solution can be viewed as a superposition of uniform density optically thin slabs
situated one behind the other, with $U$ going down from $\sim 100$ to $\sim 1$, and an optically thick slab with $U\lesssim 0.1$ on the back side. 
Therefore, RPC implies that there is a range of $U$ at a given distance, as suggested by LOC, 
but in RPC the distribution in $U$ can be calculated, rather than being a free parameter.
Similarly, $n$ is set by $\lion$ and $r$ in RPC, rather than being a free parameter.

\subsection{LINERs}\label{sec: liners}

The background of Fig. \ref{fig: BPT}, taken from K06, shows the
spread of BPT ratios in SDSS galaxies where the emission lines are excited by
a hard spectrum (above the red lines).
K06 showed that these BPT ratios have a bi-modal
distribution, indicating the existence of two distinct groups, known as
Seyferts and LINERs. K06 used G04 models with $U_0<0.03$, which are effectively constant-$n$ models, 
to show that the different narrow line ratios imply
a different $U$, where $U\sim 10^{-3}$ in LINERs, compared to $U\sim
10^{-2} - 10^{-2.5}$ in Seyferts, confirming earlier results by \cite{FerlandNetzer83}. 
Moreover, LINERs have been found to be distinct from Seyferts also in $\lledd$
(K06; fig. 9 in \citealt{Antonucci12};
\citealt{SternLaor13}), where \citeauthor{Antonucci12} and
\citeauthor{SternLaor13} found a transition $\lledd$ of $10^{-3}$.

Since a transition in the accretion flow is theoretically expected at low
$\lledd$ (\citealt{Abramowicz+95}; \citealt{NarayanYi95}), the different $U$
are thought to be a result of the different incident SED
(K06; \citealt{Ho08}).
However, why a lower $U$ follows from a different incident SED has not been
explained.
RPC may provide the missing link between $U$ and $\beta\hnu$, quantitatively.
Eq. \ref{eq: Uf} shows that a factor of $3-10$ difference in $U_{\rm f}$
implies a factor of $3-10$ difference in $\beta\hnu$. Hence, if LINERs are RPC, then either the ionizing
spectrum is harder in LINERs (larger $\hnu$), or the ratio of optical to ionizing photons is higher (larger $\beta$), or both. 
The exact difference in $\beta\hnu$ requires RPC modeling of LINERs with their observed SED.

\subsection{$\mbh$ estimates}\label{sec: mbh measurement}

The coefficient of the $\dv\propto\ncrit^{1/4}$ relation (eq. \ref{eq: FWHM vs
ncrit}) depends on $\mbh$. Therefore, this relation can be used to estimate
$\mbh$ using the $\dv$ and $\ncrit$ of the forbidden emission lines. From eq. \ref{eq: FWHM vs ncrit} we get
\begin{equation}\label{eq: mbh estimate}
M_{\rm BH} = 2\times10^7 ~\dv_{300}^2~(\beta L_{\rm i, 43})^{1/2}~n_{\rm crit,6}^{-1/2}~ \msun 
\end{equation}
where $\dv = 300~\dv_{300}\kms$, and we explicitly noted the dependence on $\beta$, which may be higher in LINERs than in Seyferts (see previous section). Eq. \ref{eq: mbh estimate} can be used to estimate $\mbh$ from each forbidden line which is emitted from $r<<\rgi$, i.e. all lines with $n_{\rm crit,6} >> 160~\lledd$ (eq. \ref{eq: rgi pressure}). 
Therefore, this method for estimating $\mbh$ is most effective in AGN with low $\lledd$, where a larger fraction of the narrow line region enters the sphere of influence of the black hole. 

For each object in Fig. \ref{fig: LINERs}, we find the $\mbh$ which best-fits the observed $\dv$ vs. $\ncrit$, for all lines with $n_{\rm crit,6}>160~\lledd$. Note that the lines which enter the fit depend on $\mbh$. We use the values of $\lion$ listed in Table \ref{tab: liners} and $\beta=2$. 
These estimates of $\mbh$ are listed in Table \ref{tab: liners}. 

In low $\lledd$ AGN the host is clearly detectable by selection. One can
therefore explore the relation of the directly measured
$\mbh$, based on gas dynamics within the black hole sphere of influence, with
various host properties, such as the bulge
mass, velocity dispersion, etc'.

\subsection{The covering factor}
\renewcommand{\fion}{f_{\rm ion}}
\newcommand{\Onodust}{\Omega({\fion=1})}

In dust-less gas with NLR densities, the emitted \Ha\ flux is determined by
the flux of incident ionizing photons. Therefore, $\lline(\Ha)/\lion$ is a
measure of  $\Omega$.
In dusty gas, one needs to correct for the absorption of ionizing photons by
dust grains.
For a typical dust distribution and AGN SED, dust dominates the opacity of
ionizing photons at $U>0.006$ (\citealt{NetzerLaor93}). In RPC most of the
absorption occurs at $U>U_{\rm f} = 0.03$ (eq. \ref{eq: Uf}), so the
fraction of ionizing photons which ionize the gas $\fion$ is expected to be
$<<1$. Indeed, in the RPC models  with $\eta=0$ and $\aeuv=-1.6$ we find
$\fion=1/7,\ 1/9,$ and $1/12$ for $Z/\zsun=1,2$ and $4$, respectively. Values
of $\aeuv$ between $-1.2$ and $-1.8$ change $\fion$ by $\pm 20\%$.

\cite{SternLaor12b} showed that if the absorption of ionizing photons by
grains is neglected, $\Onodust=0.04$ at $\lbol=10^{45.5}\ergs$. For a typical
$Z=2\zsun$, this value of $\Onodust$ implies $\Omega=9\Onodust=0.36$.
However, \citeauthor{SternLaor12b} also found that $\Onodust \propto
\lbol^{-0.3}$, reaching $\Onodust=0.4$ at $\lbol=10^{42.5}\ergs$.
Using the $\fion$ derived above we will find an unphysical $\Omega>1$ at low
$\lbol$. Hence, either $\fion$ is underestimated at all $\lbol$, or $\fion$ increases with
decreasing $\lbol$.

The true $\fion$ might be somewhat higher than we derived due to the dust
destruction mechanisms described in \S\ref{sec: dust destruction}, which we
do not model.
Also, $\fion$ may increase with decreasing $\lbol$, due to two reasons.
First, lower $\lbol$ AGN likely reside in host galaxies with lower $M_*$ and
therefore lower $Z$, which implies a higher $\fion$.
Second, an increase of $\beta\hnu$ with decreasing $\lbol$ will cause $U_{\rm
f}$ to decrease (see \S\ref{sec: liners}) and hence increase $\fion$.

\section{Conclusions}

Radiation Pressure Confinement is inevitable for a hydrostatic solution of
ionized gas, since the transfer of energy from the radiation to the gas is
always associated with momentum transfer.
Only confinement mechanisms which are stronger than RPC, or non-hydrostatic
conditions, can obviate RPC. 
The success of RPC in reproducing the observations (\S\ref{sec: observations})
suggests that these alternatives are not dominant in AGN. 

We expand on the earlier study of D02 and G04 of the RPC solution for the NLR, and
study the global structure of the photoionized gas on scales outside the BLR. 
RPC implies the following:
\begin{enumerate}
\item The value of $n$ is determined by $\lion$ and $r$, via eq. \ref{eq: n vs
r}. This relation is observed in resolved observations of the NLR, in the
FWHM vs. $\ncrit$ relation first observed by \cite{FilippenkoHalpern84}, and
in the comparison of $n_{\rm BLR}$ with $r_{\rm BLR}$ (Paper II). Together, these
observations span a dynamical range of $\sim10^4$ in $r$, from sub-pc scale
to $\kpc$ scale, and a range of $\sim10^8$ in $n$, from $10^3$ to
$10^{11}\cm^{-3}$.   

\item The hydrostatic solution of RPC gas is independent of the boundary
value $U_0$, $n_0$ or $\pgasz$. Therefore, if $r$ is known, RPC models have
essentially zero free parameters.

\item The ionization structure of RPC slabs is unique, including a highly
ionized X-ray emitting surface, an intermediate layer which emits coronal
lines, and a lower ionization inner layer which emits optical lines. The
fraction of radiation energy absorbed in each ionization state is given by
eq. \ref{eq: dtau to dlog xi}.
 This structure can explain the overlap of the extended X-ray and narrow line
emission, and some spatially resolved narrow line ratios. 

\item Beyond the sublimation radius, the line emitting gas is likely to be
dusty, at least in the layers which emit the \oiii\ and lower ionization
lines. The dust thermal IR emission constitutes $\sim 80$\% of the total
emission, and line emission the remaining $\sim 20$\%, at all $r$. Therefore,
RPC implies that there is no distinction between an IR-emitting torus, 
and a line-emitting NLR. 

\item The value of $U_{\rm f}$, which is the typical ionization parameter
where most of the radiation is absorbed, is set solely by the SED of the
incident spectrum. Therefore LINERs, which have a lower $U_{\rm f}$ than
Seyferts, are expected to have a higher $\beta\hnu$.
 \item Following G04 and K06, we find that BPT ratios implied by
RPC are consistent with observations of SDSS galaxies, assuming a nearly
constant covering factor per unit $\log r$. This covering factor distribution
is expected from the flat IR slope observed in AGN.
 \item RPC predicts that FWHM$\propto \ncrit^{1/4}$ for $\ncrit>1.6 \times 10^{8}~\lledd \cm^{-3}$, 
and implies that the normalization of the
FWHM$\propto\ncrit^{1/4}$ relation (eq. \ref{eq: FWHM vs ncrit}) depends on
$\mbh$. Therefore, $\mbh$ can be estimated directly by measuring the gas
dynamics. This method is effective in low $\lledd$ AGN, where many forbidden
lines are emitted inside the radius of influence of the black hole. In such
objects the host is expected to be well resolved, and one can therefore
explore the relation between the host properties and the $\mbh$ derived from the gas dynamics.
 \end{enumerate}

\section*{Acknowledgements}

We thank Jonelle Walsh for providing her \HST\ observations of the NLR in nearby
AGN.
Numerical calculations were performed with version C10.00 of Cloudy, last
described by \cite{Ferland+98}.
This publication makes use of data products from the SDSS project, funded by
the Alfred P. Sloan Foundation.

\appendix

\section{Ionization structure}\label{app: ionization structure}

\begin{figure*}
\includegraphics{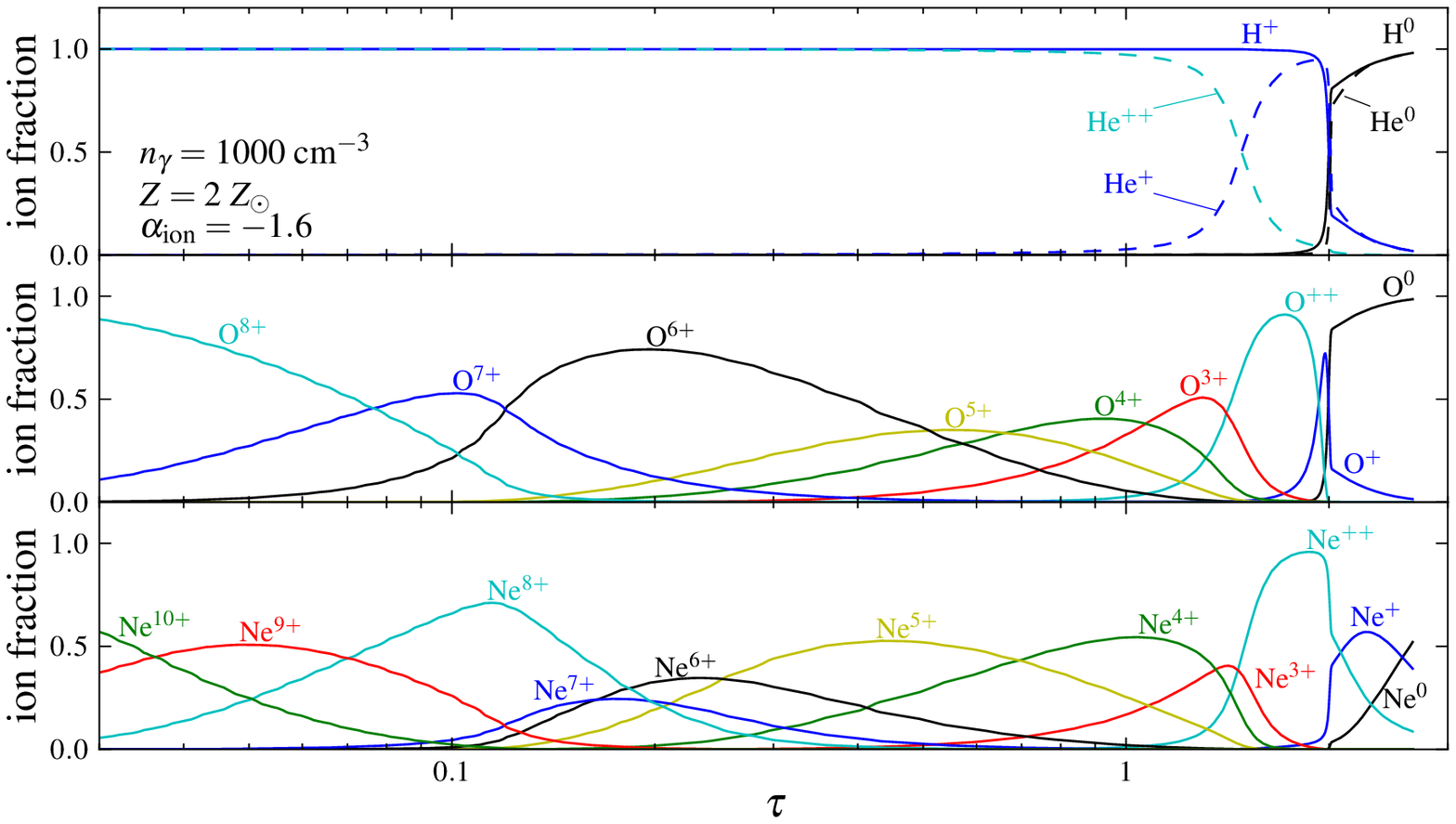}
\caption{The H, He, O, and Ne ionization structure vs. optical depth, for the RPC model shown in Fig. \ref{fig: profile vs tau}.
The broad range of ionization states implies that the same slab emits X-ray lines, `coronal' lines and low ionization optical lines.  Other reasonable choices of $\ng$, $Z$ or $\aeuv$ have only a secondary effect on the implied ionization structure. 
}
\label{fig: ionization structure}
\end{figure*}

Figure \ref{fig: ionization structure} shows the ionization structure vs. $\tau$, for the dusty RPC model shown in Fig. \ref{fig: profile vs tau}. 
Note the large range of ionization, which implies that X-ray line emission (e.g. O$^{7+}$), `coronal' line emission (e.g. Ne$^{4+}$), and \oiii\ emission can come from the same slab. 
Because the ionization structure is set to first order by $\tau$ (eq. \ref{eq: xi vs tau}), other reasonable choices of $\ng$, $Z$ or $\aeuv$ have only a secondary effect on the implied ionization structure. 

\section{Estimates of $\mbh$, $\lion$ and $\sV$}\label{app: low lledd measurements}

The references for the $\sV$ and $\mbh$ estimates used in Fig. \ref{fig: LINERs} are listed in Table \ref{tab: liner measurements}. In case of multiple measurements of the same quantity, we use the average value. In M~81 the estimate of $\mbh$ is based on gas and stellar dynamics. In other objects, if a measurement of $\sV$ is available we use the \cite{Gultekin+09} relation to estimate $\mbh$. Otherwise, we use $\mbh$ estimates based on the bulge luminosity (PKS~1718-649) or the virial assumption of the BLR (MR~2251-178 and PG~2251+113).

To estimate $\lion$ in all objects consistently, we use the measured flux of either the broad \Ha\ or broad \Hb, which are available for seven of the eight objects in Fig. \ref{fig: LINERs}. The Balmer lines are preferred also because they are less sensitive to reddening and stellar contamination effects than UV continuum observations. 
We assume $\Ha/\Hb=3$, a FRW cosmology with $\Omega=0.3,\ \Lambda = 0.7$ and $H_0 = 70\ \kms~{\rm Mpc}^{-1}$, and $\lion/\lbha=45$, appropriate for our assumed SED and the bolometric correction factor for $\lbha$ found by \cite{SternLaor12a}. For simplicity, we disregard possible changes in $\lion/\lbha$ with $\lbol$. 
If available, we use the Balmer line flux measurements from the same spectrum which was used to measure the $v$ of the NLR (see references in Table \ref{tab: liners}). In  NGC~3783 and Ark~120 the Balmer fluxes are not available in these spectra, so the broad \Hb\ flux is taken from \cite{Marziani+03}. In PKS~1718-649, the broad Balmer lines are not detected. We therefore use the narrow \Ha\ flux measured by \cite{Filippenko85}, together with the \cite{Laor03} conversion between $\lnha$ and $\lbol$. The Balmer line used in each object is listed in Table \ref{tab: liner measurements}.

To verify our $\lion$ estimates, we estimate $\lion$ also from \HST\ measurements of continuum luminosities. \cite{Sulentic+07} lists the \HST\ continuum flux at 1550\AA\ of Pictor~A, NGC~3783, Ark~120, MR~2251-178 and PG~2251+113. For NGC~7213 and M~81 we use the V-band and B-band \HST\ measurements from \cite{Lauer+95} and \cite{HoPeng01}, respectively. In PKS~1718-649, only an upper limit on the continuum flux at 1200\AA\ is available (from \citealt{KeelWindhorst91}, using \IUE). We use the bolometric correction factors from \cite{Richards+06}, again disregarding possible changes in the SED with $\lbol$. The continuum-based $\lion$ estimates of the seven objects with detections are consistent with the Balmer-based $\lion$ estimates to within a factor of 2.2 (Table \ref{tab: liner measurements}). A factor of two error in $\lion$ will change the expected $v$ in the $v$ vs. $\ncrit$ relation (eq. \ref{eq: FWHM vs ncrit}) by 20\%, significantly less than the factor of $3^{1/2}$ uncertainty in 
$v$ due to the assumed uncertainty in $\mbh$. 

\begin{table}
\begin{tabular}{l|c|c|c|c}
   & Ref. for & Ref. for & Balmer & \\
Object & $\sV$ & $\mbh$ &  line & $\frac{\lion({\rm continuum})}{\lion({\rm Balmer})}$ \\
\hline
M~81         & 1      & 6,7    & broad \Ha         & 0.6 \\
PKS~1718-649 & $^{a}$ & 8      & narrow \Ha$^{b}$  & 3$^{d}$   \\
NGC~7213     & 2      & $^{a}$ & broad \Ha         & 1.4 \\
Pictor~A     & 3      & $^{a}$ & broad \Ha$^{b,c}$ & 0.5 \\
NGC~3783     & 4, 5   & $^{a}$ & broad \Hb         & 2.2 \\
Ark~120      & 4      & $^{a}$ & broad \Hb         & 0.8 \\
MR~2251-178  & $^{a}$ & 9,10   & broad \Hb         & 0.5 \\
PG~2251+113  & $^{a}$ & 11,12  & broad \Hb         & 1.1 \\
\end{tabular}
\caption{\newline
Notes: \newline
$^{a}$ The value of $\sV$ is based on $\mbh$, or vice-versa, using the G{\"u}ltekin et al. (2009) relation. \newline
$^{b}$ Measurements were taken in non-photometric conditions. \newline
$^{c}$ We assume the broad \Ha\ flux is 70\% of the quoted total (broad+narrow) \Ha\ flux. \newline
$^{d}$ Continuum-based measurement is an upper limit. \newline
References: 
1. G{\"u}ltekin et al. (2009); 2. Nelson \& Whittle (1995); 3. Lewis \& Eracleous (2006); 4. Onken et al. (2004); 5. Garcia-Rissmann et al. (2005); 6. Bower et al. (2000); 7. Devereux et al. (2003); 8. Willett et al. (2010); 9. Zhou \& Wang (2005); 10. Kelly \& Bechtold (2007); 11. Vestergaard \& Peterson (2006); 12. Davis \& Laor (2011).
}
\label{tab: liner measurements}
\end{table}

\section{Emission line ratios}\label{app: emission line ratios}

The observed emission line ratios which appear in Fig. \ref{fig: U dependence} are assembled as follows. 
The observed optical \nev, \oiii\ and $\oi$ luminosities are taken from the \cite{SternLaor12a} sample of $z<0.3$ broad line AGN, which was selected from the 7\th\ data release of the SDSS (\citealt{Abazajian+09}) based on the detection of broad \Ha\ emission. The $\oiii$ detection rate is 99.7\%. 
In the $z>0.15$ objects where $\nev$ enters the SDSS spectrum, the $\nev$ detection rate is 96\%. 
We use the \citeauthor{SternLaor12a} \oiii\ measurements, and the SDSS pipeline measurements of \nev. 
We find $\log \nev/\oiii = -0.79\pm0.4$ (one-sigma range) in the 973 $z>0.15$ objects, and $\log \nev/\oiii = -0.95\pm0.4$ in the 172 objects with $L_{{\rm i,} 45}>1$. 
For comparison, \cite{Nagao+03} find similar $\log \nev/\oiii = 0.82\pm0.32$ on 34 local AGN, \cite{Zakamska+03} find a mean $\log\ \nev/\oiii=-0.92$ in type 2 quasars, and \cite{VandenBerk+01} find a mean value of $-0.8$ in SDSS quasars. 
The \oi\ emission line is detected in 76\% of the sample. We find $\log \oiii/\oi = 1.1\pm0.3$, taking upper limits in objects with no detections. The mean $\oiii\ /\ \oi$ in the $L_{{\rm i,} 45}>1$ objects is $1.2$, though the \oi\ detection rate is 55\%. \cite{VandenBerk+01} and \cite{Zakamska+03} find a similar mean $\log \oiii/\oi=1.1$ and $1.0$, respectively.

The IR \neiii, \nev, and \nevi\ fluxes are taken from the 36 PGs in the QUEST (\citealt{Veilleux+09}) sample. We find $\log\ \nev/\neiii=-0.3\pm0.2$ in the 35 objects with a \neiii\ detection. In the eight objects without a \nev\ detection, the $3\sigma$ upper limits are used. 
Similar values of $\nev/\neiii$ were found by \cite{Gorjian+07}, in AGN which span four orders of magnitude in \nev\ luminosity. 
Also, \cite{Weaver+10} found a similar $\log \nev/\neiii \approx -0.4$ in 130 AGN with lower luminosities than the PGs, selected by their hard X-ray emission.
In seven of the 35 PGs, $\nevi$ was observed, and detected in all objects. We find $\log\ \nevi/\neiii = -0.06\pm0.25$ in these objects. 

\end{document}